\documentclass[preprint,12pt]{elsarticle}

\usepackage[T1]{fontenc}
\usepackage{lineno}
\usepackage{fdsymbol}
\usepackage{amsmath,amsfonts}
\usepackage{siunitx}
\usepackage{algorithmic}
\usepackage{tablefootnote}
\usepackage{soul, color}
\usepackage{ dsfont }
\usepackage{amsmath}

\usepackage{array}
\usepackage{makecell}

\usepackage{booktabs}

\usepackage{caption}
\usepackage{subcaption}

\usepackage{multirow} 
\usepackage[title]{appendix}

\usepackage{tabularray}
\usepackage{xcolor}

\journal{arxiv}

\begin{document}

\begin{frontmatter}




\title{Extra Throughput versus Days Lost in load-shifting V2G services: Influence of dominant degradation mechanism}


\author[inst1]{Hamidreza Movahedi}
\author[inst1]{Sravan Pannala}
\author[inst1]{Jason Siegel}
\author[inst2]{Stephen J. Harris}
\author[inst3]{David Howey}
\author[inst1]{Anna Stefanopoulou}

\affiliation[inst1]{organization={Department of Mechanical Engineering, University of Michigan},
            addressline={1231 Beal Ave}, 
            city={Ann Arbor},
            postcode={48109}, 
            state={MI},
            country={USA}}
\affiliation[inst2]{organization={Energy Storage and Distributed Resources Division, Lawrence Berkeley National Laboratory},
            city={Berkeley},
            postcode={94720}, 
            state={CA},
            country={USA}}
\affiliation[inst3]{organization={Department of Engineering Science, University of Oxford},
            addressline={Parks Road, OX1 3PJ}, 
            city={Oxford},
            country={United Kingdom}}

\begin{abstract}

Electric vehicle (EV) batteries are often underutilized. Vehicle-to-grid (V2G) services can tap into this unused potential, but increased battery usage may lead to more degradation and shorter battery life.
This paper substantiates the advantages of providing load-shifting V2G services when the battery is aging, primarily due to calendar aging mechanisms (active degradation mechanisms while the battery is not used). 
After parameterizing a physics-based digital-twin for three different dominant degradation patterns within the same chemistry (NMC),
we introduce a novel metric for evaluating the benefit and associated harm of V2G services: \textit{throughput gained versus days lost (TvD)} and show its strong relationship to the ratio of loss of lithium inventory (LLI) due to calendar aging to the total LLI
($\text{LLI}_\text{Cal}/\text{LLI}$).
Our results that focus systematically on degradation mechanisms via lifetime simulation of digital-twins significantly expand prior work that was
primarily concentrating on quantifying and reducing the degradation of specific cells by probing their usage and charging patterns. Examining various cell chemistries and conditions enables us to take a broader view and determine whether a particular battery pack is appropriate for load-shifting (V2G) services.
Our research demonstrates that the decision "to V2G or not to V2G" can be made by merely estimating the portion of capacity deterioration caused by calendar aging. Specifically, TvD is primarily influenced by the chemistry of cells and the environmental temperature where the car is parked, while the usage intensity and charging patterns of EVs play a lesser role.

\end{abstract}

\begin{graphicalabstract}
\includegraphics{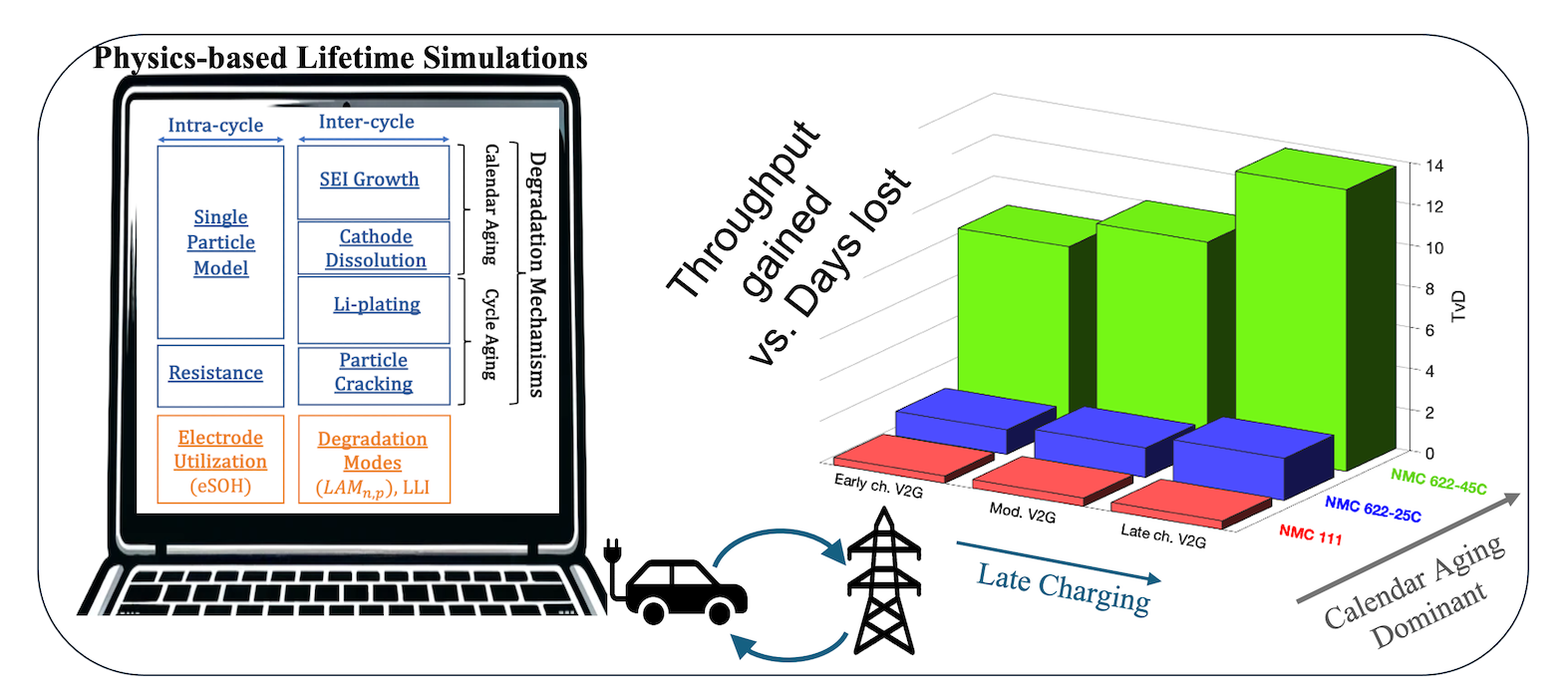}
\end{graphicalabstract}

\begin{highlights}
\item Simulation of lifetime degradation due to V2G using physics-based EV battery digital-twins of three cell families with distinct dominant degradation mechanisms. 
\item Introduction of a new metric for quantifying the degradation cost and V2G benefits: \textit{throughput gained versus days lost (TvD)} ratio of V2G services, where the \textit{throughput gained} is the normalized additional battery utilization in Ah throughput, while the \textit{days lost} is the relative lost lifespan of the battery due to V2G usage.
\item Offering physics-based justification and qualification of the popular belief: "Use it or lose it". If calendar aging is more significant than other cycling aging mechanisms, we might as well use the battery for V2G.
\item Evaluation of the secondary impact of charging protocol timing and driving distance on battery degradation in the presence of V2G services.
\end{highlights}

\end{frontmatter}


\section{Introduction}
\label{sec:Introduction}

The heightened importance of the climate crisis and the growing integration of renewable energy sources in the grid are making electric vehicles (EVs) more attractive for vehicle-to-grid (V2G) services, where EV batteries are used as extra energy storage to support the grid. 
These services, which are becoming an exciting subject in research and industry\ \cite{inci2022integrating,collath2022aging,pearre2019review}, can be categorized into load-shifting V2G and ancillary services\ \cite{ma2016overview}. 
Load-shifting services can reduce peak power demand on the grid. Upgraded EVs act as controllable grid storage by discharging when needed, reducing reliance on fossil fuels. These services are also beneficial for integrating renewable energy sources such as solar into the grid since solar generation has limited operational hours that may not align with peak industrial or building demand periods\ \cite{kempton2005vehicle}.

Passenger light-duty vehicles are often parked and not driven, resulting in underutilized batteries for EVs\ \cite{etxandi2023electric}. V2G technology, while supporting the grid, also provides financial profit for owners, especially considering that batteries degrade even when stored and not used.
This compensation is particularly relevant for car owners who drive short distances and may reach the time limit of their pack warranty before hitting the total mileage and current throughput (Ah) limit. Such under-utilization allows car owners in regions with high electricity tariffs to leverage their battery pack for financial gain.
In addition, Li-ion battery materials are limited natural resources in high demand\ \cite{iea2021role}, and battery manufacturing contributes significantly to EV carbon footprint\ \cite{kawamoto2019estimation}. Therefore, efficient battery use is crucial, and the environmental benefits of V2G services are two-fold: grid decarbonization and efficient use of natural resources.


On the other hand, the usage of EV batteries for additional services will result in higher degradation.
EV manufacturers may account for the participation in V2G by calculating the "virtual miles" toward the warranty limit \cite{UNGTR22}.
However, the definition of virtual miles fails to differentiate between the various causes of degradation and does not account for the fact that the battery pack will degrade even when the EV is not in use and parked.
Given that battery replacement cost governed by the warranty could significantly impact EVs' overall cost of ownership \ \cite{konig2021overview}, the extra degradation due to V2G might be a significant deterrent for EV owners. Therefore, a metric that reflects the additional degradation caused exclusively by the V2G d services after accounting for the degradation that would have occurred anyway if the pack was not cycled is required.

To this end, we employ a digital-twin as a tool to explore different V2G scenarios and their associated increased battery degradation.
By introducing the \textit{throughput gained versus days lost ratio (TvD)} and using a physics-based model for the V2G digital-twin, we analyze the sensitivity of the TvD to battery degradation mechanisms, and other factors such as depth of discharge, driving distance, and average state of charge (SOC) of battery duty cycles.  
Furthermore, our work clarifies the mixed results reported in previously published works on battery degradation under V2G\ \cite{peterson2010lithium, Kim2022, guo2019impact, UDDIN2018342}, giving a better understanding of the observed degradation patterns.
Especially since our physics-based model is parameterized using experimental data from cells with various degradation modes, allowing for reliably studying a wide range of scenarios.

Previous studies on battery degradation due to load-shifting V2G services show conflicting results, as summarized in Table \ref{Table: past works}. 
Some studies suggest that V2G services can be very detrimental to battery longevity \cite{Dubarry2017, wang2016quantifying}. In contrast, others have shown a more optimistic prognosis and claim that such services can even increase battery life\ \cite{wei2022comprehensive,uddin2017possibility}.
These contrasting results are attributed to variations in chemistries, temperature conditions, and intensity of the V2G services\ \cite{guo2019impact}. 
Our digital-twins explain these discrepancies using physics-based simulations and clarify the degradation during the V2G and noV2G duty cycles.
We distinguish between the contribution of effective degradation modes and mechanisms. 
We show that the fractional loss of lithium inventory ($\text{LLI}$) specifically caused by calendar aging ($\text{LLI}_\text{Cal}$) is the deciding factor for determining the degradation of cells used for load-shifting V2G services, with minor influence from driving, charging, and V2G depth of discharge.
This clarification is essential for reconciling these discrepancies.

\begin{table}[!htb]
    \centering
    \tiny
        \begin{tabular}{p{{5em}}>{\centering\arraybackslash}p{45pt}>{\centering\arraybackslash}p{7pt}>{\centering\arraybackslash}p{5pt}>{\centering\arraybackslash}p{55pt}>{\raggedright\arraybackslash}p{165pt}} 
         \toprule
       \makecell{Paper \\ source}& Cell type&Exp. \ \ \ \ \ & Model& \makecell{Duty cycle \\factors}  &  \makecell[c]{Conclusions}\\  \hline

     \makecell[l]{Bhoir \\ et al.\cite{bhoir2021impact} }& NMC cells & N & Emp. & \makecell{Intensity\\ DOD}  & $+$ Peak shaving reduces degradation.\\
     
     \makecell[l]{Fioriti \\ et al.\cite{fioriti2023battery} }& \makecell{Pouch NMC\\ 20Ah cells} & N & Emp. & \makecell{Driving dist.\\ Temp.\\ Charg. Patt.}  & \makecell[l]{\textbullet \ 0.1\% increase in Ah throughput\\ \textbullet \ 3.8\% faster degradation}\\

     \makecell[l]{Wei \\ et al. \cite{wei2022comprehensive} }& \makecell{Pouch NMC\\ 24Ah cells} &N &Semi emp & \makecell{ Intensity \\ DOD}& \makecell[l]{\textbullet \ Average SOC main factor\\ $+$ \ V2G could even slow the degradation by \\ \ \ \ slowing calendar aging.}\\ 
     
     \makecell[l]{Bishop \\ et al.\cite{bishop2013evaluating} }& \makecell{Cylind. LFP\\ 2.2Ah cells}& N& Emp.& \makecell{Driving dist.\\ DOD\\ Charg. patt.} & \makecell[l]{\textbullet \ Degradation mostly a function of energy throughput\\
    \textbullet \ Sequence of resting and charging is also influential}\\

      \makecell[l]{Jafari \\ et al.\cite{jafari2017electric}} &\makecell{ Cylind.LFP\\ 2.4Ah cells} & N & Emp. & \makecell{Driving dist.\\ Temp.\\ Charg. Patt.}  & \makecell[l]{ $-$ Peak shaving does not change Ah throughput \\ \ \ \ but leads to 37\% more degradation }\\

     \makecell[l]{Peterson \\ et al.\cite{peterson2010lithium}}& \makecell{Cylind. LFP\\ 2.2Ah cells}&   Y&    -&  
     \makecell{Intensity\\DOD}
     &\textbullet \ Degradation is a function of energy throughput\\  

     \makecell[l]{Zheng\\ et al.\cite{zheng2022economic}} & LFP pack & N & Emp. & \makecell{Charg. Patt.\\ Discharg. time}  & \makecell[l]{\textbullet \ Battery wear is related to energy throughput \\ $-$ \ 
     Highly likely for EV aggregators to operate at a \\ \ \ \ loss for current battery costs.}\\

     \makecell[l]{Zheng\\ et al.\cite{zheng2023modeling}} & LFP pack & N & Emp. & \makecell{Charg. Patt.\\ DOD}  & \makecell[l]{\makecell[l]{\textbullet \ Battery degradation assumed to be a function of \\ \ \ \ charging rate \\ $-$ \ 
     Not economically worth it to sell to the grid \\ \ \ \ due to degradation.}}\\

     \makecell{Dubarry et \\ al.\cite{Dubarry2017, Dubarry2018} }& \makecell{Cylind. NCA\\ 3.35Ah cells}& Y& -& \makecell{ Sequence \\Charg. patt.}& \makecell[l]{\textbullet \ Increase in LAM results in worsened degradation.\\ $-$ V2G operations have a significant
     detrimental effect. }\\

     \makecell[l]{Gong \\ et al. \cite{GONG2024100316} }& \makecell{Cylind. NCA\\ 3.35Ah cells}& Y & Emp.& \makecell{Diff. V2G\\ Charg. patt.} & \makecell[l]{ $+$ Minimal difference between strategies\\ \ \ \ could even be attributed to cell-to-cell variance}  \\

     \makecell[l]{Uddin \\  et al.\cite{uddin2017possibility} }
    & \makecell{Cylind. NCA\\3.0Ah cells}& N& Emp.& \makecell{Intensity \\ Driving dist.}&$+$ \ V2G can be optimized to reduce degradation.\\

     \makecell[l]{Kim \\ et al. \cite{Kim2022}}& \makecell{Cylind.cells \\ LFP:2.5Ah \\ NCA:3.4Ah,\\  NMC:3.5Ah\\ NC:3.0Ah }&Y &- & \makecell{Diff. grid \\duty cycles} & \makecell[l]{\textbullet \ Degradation is mainly due to  SOC and  ($\Delta \text{SOC}$) \\ \textbullet \ Effect of V2G varies substantially among different \\ \ \ \ chemistries}\\

     \makecell[l]{Petit \\  et al.\cite{petit2016development}} & \makecell{ Cylind.cells\\LFP:2.3Ah\\ NCA:7Ah  }& N& Emp.& \makecell{Intensity\\ Charg. patt.}&\makecell[l]{\textbullet \ NCA  mainly degrade due to charge throughput \\  \textbullet \ LFP cells degrade due to average SOC}\\ 
       
     \makecell[l]{Wang\\ et al.\cite{wang2016quantifying}} & \makecell{Cylind. \\NMC/NMO\\1.5Ah cells} & N & Semi emp & \makecell{Intensity\\ Temp.\\ Driving dist.}  &\makecell[l]{$-$ \ Regular peak shaving  can cut the battery life \\ \ \ \ by half.}\\

     EPRI \ \cite{EPRI}& \makecell{Chrys. Paci.\\ PHEV pack}& Y & - & - & \makecell[l]{\textbullet \ V2G led to 165\% more capacity degradation \\ \ \ \ yielded 70\% more Ah throughput.}  \\

    \makecell[l]{ Thingvad \\ et al.\cite{Thingvad2021} }
    & \makecell{Nissan \\e-NV200}& Y& Emp.& - & \textbullet \ Did not compare with a baseline (noV2G)\\

    \bottomrule
        \end{tabular}
      \caption{Previous studies on V2G degradation}\label{Table: past works}
\end{table}

There is also a lack of consensus among studies on the factors that affect battery degradation. 
Specifically, some of the previous V2G studies have reported that battery degradation is primarily proportional to SOC levels and the test duration in time \cite{uddin2017possibility, Kim2022}. In these studies, V2G services were shown to mitigate the capacity fade compared to a noV2G scenario for the same amount of Ah throughput. The cells used are most likely prone to calendar aging mechanisms, so processing more Ah throughput via V2G would not significantly impact the total degradation. 
However, other studies have reported a high correlation between Ah throughput and capacity fade \cite{peterson2010lithium, Dubarry2017}. 
Unlike the former cases, cells used in these latter studies are especially prone to degradation mechanisms that get amplified due to additional Ah throughput from V2G, such as particle cracking, as the primary source of degradation. 
Here, we explain the inconsistency between these two cases (i.e., whether the limiting case is calendar aging or cycle aging) by considering various degradation mechanisms and a physics-based model tuned for different cell families representing these two opposite cases. 
We clarify that even with the same chemistry anode (graphite) and cathode (NMC), there are different degradation patterns dominated by either cycling damage (particle cracking) or calendar aging (SEI growth).

Furthermore, we demonstrate the need for separate calendar and cycle aging models to quantify the benefits and drawbacks of the V2G services accurately. Our study gives insight into the importance of considering the underlying degradation mechanisms that might trigger higher degradation or extend the Ah throughput utilization before reaching the battery pack's end-of-life (EOL).

Numerous studies\ \cite{leippi2022review} have simulated the degradation of the Li-ion batteries due to V2G services. These studies create a degradation model based on fast aging experiments and then impose V2G scenarios on the devised models. Given that many of these studies aim to optimize financial profits from V2G, they rely on empirical \cite{bhoir2021impact,tchagang2020v2b,zheng2022economic} or, at best, semi-empirical \cite{wang2016quantifying,wei2022comprehensive} models to simplify calculations.

Physics-based models have been shown to improve the degradation prediction for cells in grid-connected energy storage facilities (where driving is not included in the duty cycle).
Reiners et al.\ \cite{reniers2018improving} modeled the battery degradation using three models with different levels of complexity. 
Even though their most complex model only included solid-electrolyte interface (SEI) growth as the sole degradation mode, they experimentally demonstrated that using this model can improve battery utilization and profit substantially\ \cite{reniers2021unlocking}. They also showed how overly simplified degradation models can result in erroneous conclusions. In contrast to their work, our paper includes driving in the duty cycle and additional battery degradation mechanisms to simulate an automotive pack that provides energy storage services.

To the best of our knowledge, this is the first work to use a physics-based model to predict the degradation of Li-ion cells in an automotive pack used for regular driving and V2G operations. 
The closest study to a physic-based model is the work done by Li et al. \cite{li2023degradation}, where they only consider SEI as the degradation mechanism and match the capacity of the cell only at two points in the life of the battery and ignore the degradation profile.

In our work, we tune detailed degradation models based on a single-particle model (SPM) using experimental accelerated aging data and its whole profile to identify the degradation model parameters uniquely.
Then, we use the tuned model to simulate V2G and noV2G scenarios for three different cell families (various chemistries and conditions) with different dominant degradation mechanisms.
We calculate the contribution of each degradation mechanism to the capacity fade and relate the benefit of the V2G operation to the contributing portion of each mechanism. This work shows that the extent of each mechanism's contribution can determine the TvD of V2G services over the battery lifetime.

Contributions of this work include introducing:
\begin{itemize}
    \item A physics-based model for analyzing the intra/inter-cycle degradation during load-shifting V2G services.
    \item A range of different cell families with different dominant degradation mechanisms and use conditions reflecting an exemplary but demanding driving schedule (more than 68 miles/day).
    \item A new metric for quantifying the degradation cost and V2G benefits: \textit{throughput gained vs. days lost (TvD)} ratio of the V2G services. The gain is the relative additional throughput provided to the grid, while the days lost are the relative lost life due to increased degradation. 
    \item Analysis of the effect of late- and early-charging protocols and the driven distance per day on battery degradation during V2G services for cells with different dominant degradation modes.
    
\end{itemize}   

This study paves the way for quantifying the additional encumbered environmental and financial implications of V2G for different stakeholders, namely the car owners, the utility companies, and the workplace and the home in which V2G is operated. 
We mainly focus on discovering the untapped throughput of batteries before 70\% capacity fade is reached but implicitly show how utilizing this potential can benefit the environment. 
This environmental aspect has been largely ignored previously\ \cite{sovacool2018neglected} and will be studied in future work.

The outline of the paper is as follows. 
Section 2 provides a background on previous research and explains how our work is related to it. In Section 3, we introduce the V2G strategies and vehicle driving cycles.
Section 4 describes the considered degradation mechanisms and the model tuning process. The simulation results are presented in Section 5 and discussed in Section 6. Section 7 includes the conclusions and future work.

\section{V2G services} \label{V2G service section}
Load-shifting V2G services can reduce the height of the peak power demand by charging EV batteries during off-peak periods and discharging them to the grid during peak times\ \cite{wang2013grid}.

Here, we assume the V2G services have consistent daily operation.
This section describes three V2G service duty cycles and a noV2G baseline duty cycle used in simulations. We will also explain the drive cycle contained within these daily drives and V2G duty cycles.

\subsection{Duty cycles}
\begin{figure}[htb!]
    \centering
    \includegraphics[width=0.9\textwidth]{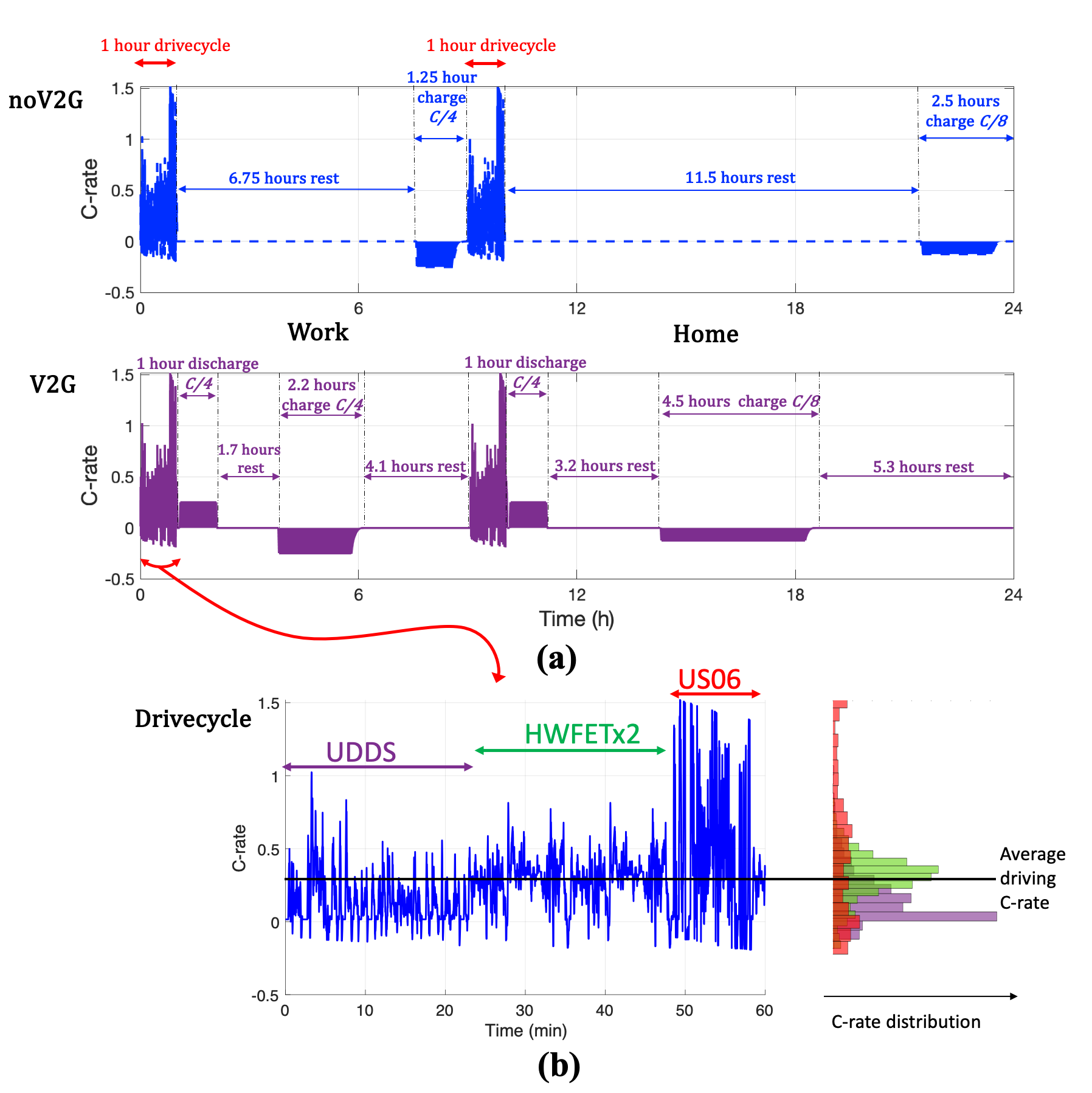}
    \caption{Scenarios in this study: (a) Daily duty cycles (24 hr) with and without V2G are considered, including one hour of driving to work and an hour back home. The V2G duty cycle includes 2 hours of discharge to the grid divided equally between morning and afternoon. The average SOC is equal to 0.79 for both duty cycles. (b) The current profile (as C-rate) of the drive cycle includes Federal test procedures with an accumulated driving distance of 34.1 miles each hour. The C-rate distribution of each drive cycle is also presented. 
    }
    \label{fig: duty cycle and drivrcycle}
\end{figure}

\begin{figure}[!ht]
    \centering
    \includegraphics[width=0.9\textwidth]{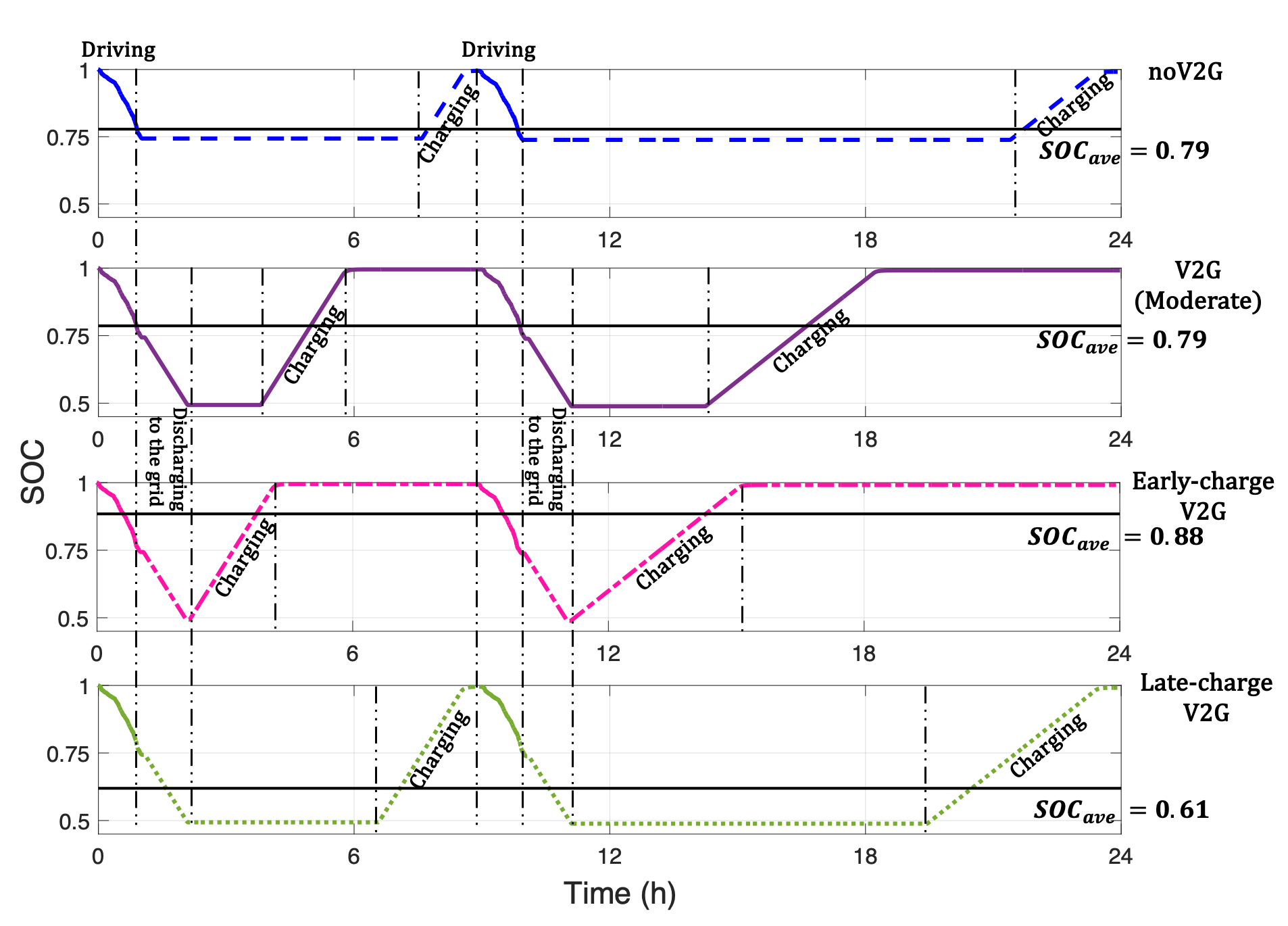}
    \caption{
    The resulting SOCs in the first cycle for the noV2G, V2G, early-charge V2G, and late-charge V2G scenarios. The average state of charge $(\text{SOC}_\text{ave})$ is shown with a solid black line for each scenario.
    The noV2G and V2G cases have the same average SOC; thus, we call this case \textbf{moderate} V2G. The early-charging V2G case rests longer at higher voltage and has a larger average SOC. Late-charging V2G rests at a lower voltage and has a smaller average SOC.}
    \label{fig: SOCs}
\end{figure}

The duty cycles we consider are roughly based on cases tested by Dubarry et al.\ \cite{Dubarry2017} as shown in Fig.\ \ref{fig: duty cycle and drivrcycle}(a). One of these duty cycles presents the least damaging case of noV2G (i.e., charging right before driving), and the second is a V2G scenario with the same average SOC level ($\text{SOC}_\text{ave}$) as the noV2G case, as illustrated in Fig.\ \ref{fig: SOCs}. 
These two scenarios, with identical daily $\text{SOC}_\text{ave}$, were selected initially to eliminate any impact of the $\text{SOC}_\text{ave}$ on degradation.

Later in the paper, we will analyze the impact of SOC by considering V2G services with different average SOCs, resembling users who choose early- or late-charging protocols after discharge to the grid. These scenarios are also presented here and shown in Fig.\ \ref{fig: SOCs}.

\begin{itemize}

\item NoV2G (baseline):
This scenario includes two hours of driving, emulating a one-hour drive to work in the morning and one hour back home, as shown in Fig.\ \ref{fig: duty cycle and drivrcycle}(a). 
The battery is charged at C/4 rate at work, right before driving back home, and C/8 at home right before driving to work in the morning. For the remainder of the noV2G duty cycle, the battery is resting. The charging happens directly before driving periods to create the least damaging noV2G scenario. For the remainder of the noV2G duty cycle, the battery rested.
This sequence reduces the resting SOC level, as shown in Fig.\ \ref{fig: SOCs}, and consequently mitigates the calendar aging caused by SEI, which is more prominent at high resting SOC values\ \cite{safari2008multimodal}. The minimum SOC in this scenario is 0.73,  and the average SOC is 0.79 in the first cycle. Needless to say, as the battery ages, the capacity of the cells fades, leading to a reduction in the minimum SOC and widening of the range of the SOC window over the lifetime simulation.

\item V2G (moderate V2G):  In this scenario, in addition to charging and driving, the electric vehicle was discharged to the grid at C/4 for one hour after each driving period to simulate load-shifting grid services. The vehicle was assumed to arrive at the workplace (after the morning drive) or home (after the afternoon drive).
The charging intervals in the morning and afternoon were chosen to match the average SOC to that of the noV2G case ($\text{SOC}_\text{ave}$=0.79).
In this scenario, the SOC decreases to 0.49 in the first cycle, widening the used SOC range $(\Delta \text{SOC})$ from 0.27 $(\text{SOC}
: 0.73-1)$ to 0.51 $(\text{SOC}: 0.49-1)$ as shown in the first and second subplots of Fig.\ \ref{fig: SOCs}.

\item  Early-charging V2G: The battery was charged right after driving in this duty cycle. Hence, the average SOC has a higher value $(\text{SOC}_\text{ave}=0.88)$ than the moderate V2G scenario.

\item Late-charging V2G: Similar to the noV2G case, the battery was charged directly before driving to lower the average state of charge $(\text{SOC}_\text{ave}=0.61)$, which is therefore below the moderate V2G case.

\end{itemize}

\subsection{Drive cycle}
For the driving part of the duty cycles, we assume a morning and an afternoon commute that lasts an hour each way. Each commute consists of one Urban Dynamometer Driving Schedule (UDDS), two Highway Fuel Economy Tests (HWFET), and a high acceleration supplemental federal test (US06)\ \cite{EPA_drivecycles}. These drive cycles were primarily introduced by the EPA to determine vehicle emissions and fuel economy and were created to represent a typical vehicle velocity profile for different driving conditions\ \cite{prakash2016use}. The required current load to the batteries was extracted from the experimental data used in Mohtat et al.\ \cite{Mohtat2021}. The resulting C-rate and its distribution are presented in Fig.\ \ref{fig: duty cycle and drivrcycle}(b).
The total driving distance of the combined drive cycle is 34.1 miles.
This is longer than the distance that most (more than 90\%) EVs drive daily\ \cite{ZHAO20232537, DONG201444}, making this driving schedule demanding and one where introducing V2G could be challenging. As this longer travel distance will degrade the battery faster than usual, the potential use of V2G may also be limited. Later in the paper, we analyze the effect of shorter drive cycles.

\section{Lifetime degradation model}
\begin{figure}[htbp]
    \centering
    \includegraphics[width=0.99\textwidth]{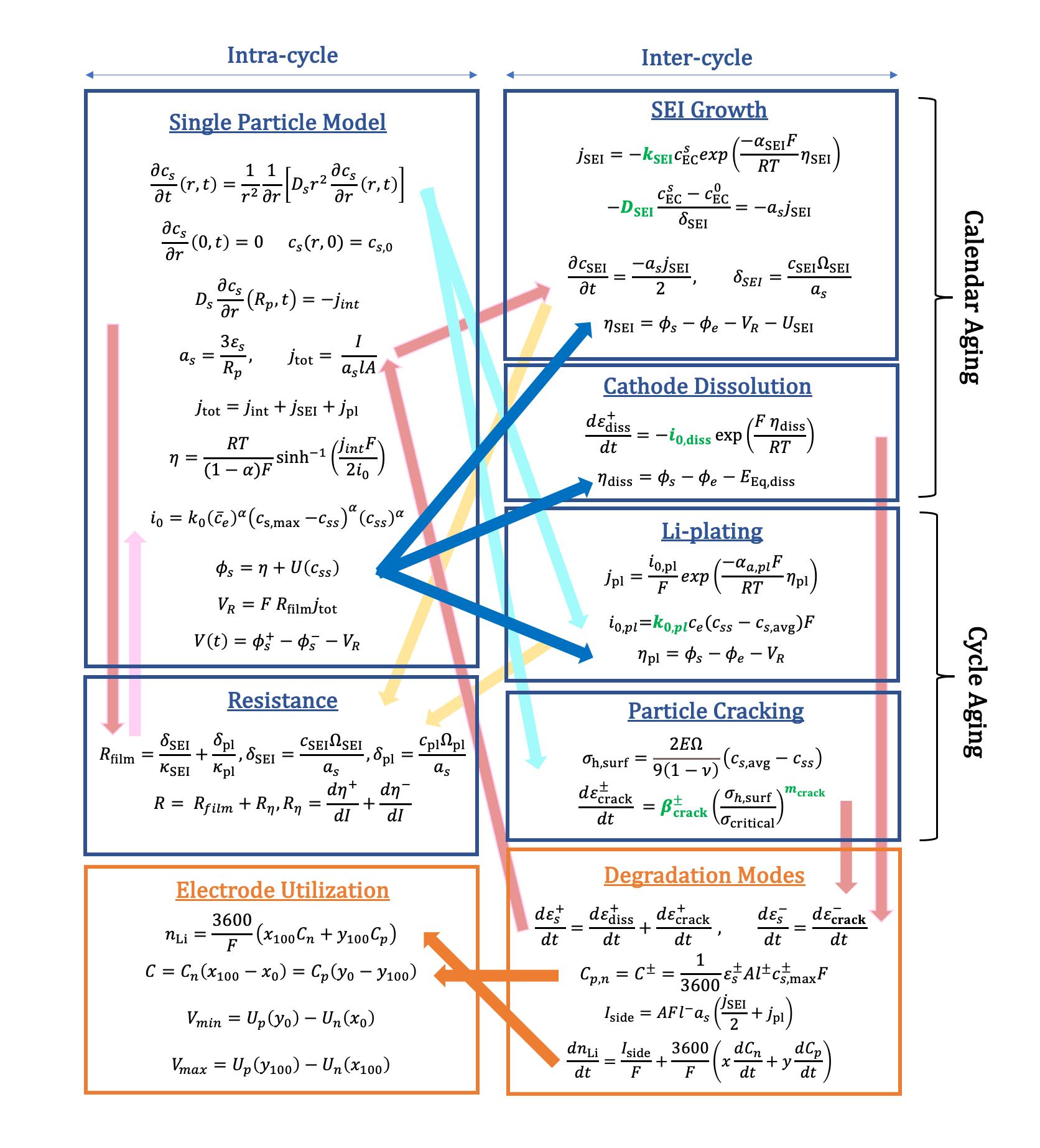}
    \caption{Summary of degradation mechanism equations based on SPM. Cathode dissolution and SEI growth are the only mechanisms that are active during calendar aging. All the mechanisms are present during the cycling of the cell.}
    \label{fig: Deg_mechanisms_eq}
\end{figure}

In this section, we summarize the predictive reduced-order electrochemical model used to simulate Li-ion aging. We briefly describe the degradation mechanisms for cell degradation of various lithium-nickel-manganese-cobalt (NMC) cell families that exhibit different dominant degradation modes. 
The digital-twin allows a comprehensive exploration of these degradation mechanisms under various cycling conditions, C-rates, Ah throughput conditions, SOC windows, and average SOCs depending on delaying charging. 
Additionally, we present the experimental data used to tune the digital-twin for three cell families with different dominant degradation mechanisms and outline the fitting procedure used to parameterize these degradation mechanisms for the lifetime degradation model.
Fig.\ \ref{fig: Deg_mechanisms_eq} provides a schematic representation of how the digital-twin operates.
The model takes current profile inputs relating to the usage pattern, simulates the physical degradation mechanisms, and computes the degradation modes associated with the loss of lithium inventory (LLI) or LAM in each electrode. 
The digital-twin includes degradation mechanisms that occur independently of battery usage or Ah throughput, even during parking (when no current is flowing), by inducing LLI due to calendar aging. 
These mechanisms correspond to the SEI growth in the anode particles and the cathode transition metal dissolution, which will be explained later in this section. The integration and interaction of all these physical degradation mechanisms for different use patterns are simulated over an entire day (intra-cycle) and computed day after day (inter-cycle) as the battery ages. A summary of the governing equations and their interconnections within intra and inter-cycles are shown in Fig.\ \ref{fig: Deg_mechanisms_eq} and are briefly described below.

\subsection{Intra-cycle electrochemical model}
The degradation depends on complex dynamic phenomena across the cell electrodes. The left upper panel of  Fig.\ \ref{fig: Deg_mechanisms_eq} summarizes the digital-twin equations using a single particle model (SPM) in which each electrode is approximated by assuming that the radii of all electrode particles are the same, and therefore a single average-radius particle adequately captures behavior. This physics-based model ignores the electrolyte dynamics and the spatial variation of reaction current yet provides enough accuracy for low to moderate C-rates\ \cite{li2018single}.
The solid-phase concentration $(c_s)$ in each particle is calculated using a partial differential equation (PDE) for diffusion, with boundary conditions at the center and surface of the sphere. The intercalation overpotential $ \eta$ is calculated using the Butler-Volmer equation. The terminal voltage of the cell is given by the difference in potential of the negative and positive electrodes $(\phi_{s}^+ - \phi_{s}^-)$, minus the losses due to ohmic resistance $(V_R)$. The characteristics of the cells are presented in Table \ref{characteristics table}, and the parameters used for the electrochemical model are given in \ref{appendix.SPM papameters}.

\begin{table}[!htb]
    \centering
    \scriptsize
        \begin{tabular}{ p{0.5em} c c c c} 
         \toprule
       Par. & Description & NMC111 & NMC622-25C & NMC622-45C \\  \hline
         -  & Anode chemistry& \makecell{Graph. (Hitachi\\ MAG-E3)}& \makecell{Graph. Sup.\\  SLC 1520-T}&\makecell{Graph. Sup.\\ SLC 1520-T} \\ 
         -    &   Cathode chemistry  & NMC111 & NMC622 & NMC622 \\     \hline
         $C$    &  \makecell{Nominal capacity (Ah)}             &  5    &  3.5  & 2.5  \\   \hline
         $A$    & \makecell{Cell surface area ($m^{2}$) }       & 0.205 & 0.135 & 0.101 \\  
         $R_s^{-}$  & \makecell{Neg. particle rad. ($\mu m$ )}  & 10    &  13.5    &  10 \\   
         $R_s^{+}$  &  \makecell{Pos. particle rad. ($\mu m$ )}  &     3.5 &  2.1    &   3.5\\    
         $l^{-}$  & \makecell{Neg. elec. thick. ($\mu m$ )}  &   62.0   &    55.7   &  61.6 \\   
         $l^{+}$  & \makecell{Pos. elec. thick. ($\mu m$ )}  &   67.0   &    55.6   &  54.5 \\  
         \bottomrule
        \end{tabular}
      \caption{Characteristics of the three NMC cell families considered.}\label{characteristics table}
\end{table}

\subsection{Inter-cycle degradation mechanisms}
There is a plethora of degradation mechanisms that can explain aging in batteries. Here, we considered four mechanisms that are most commonly assumed\ \cite{edge2021lithium}  to cause degradation during normal operation: (a) SEI growth, (b) transition-metal dissolution in the cathode, (c) mechanical degradation due to particle cracking in the electrodes, and(d) lithium plating. 

Dissolution and SEI growth are time-dependent processes. While they are active during storage, leading to calendar aging, these mechanisms also occur when the battery is in use. All four mechanisms, including SEI growth and dissolution, remain active when the battery is used.

The detailed equations of the models of these mechanisms and their interactions with SPM are shown in Fig.\ \ref{fig: Deg_mechanisms_eq}. 
The following descriptions provide an overview of these mechanisms.

\subsubsection{SEI growth} 
As a result of the reactions between the electrolyte and the negative electrode in the negative electrode, a layer of reaction products forms on the solid-electrolyte interface. These reactions consume lithium, so the capacity of the cell decreases.
Growth of the SEI layer is governed by diffusion and kinetic limiting behaviors.
This degradation mechanism is more pronounced at higher SOCs and cell temperatures\ \cite{safari2008multimodal}. The parameters in this model that will be tuned include $k_\text{SEI}$ and $D_\text{SEI}$ which are the kinetic rate constant and the diffusivity of the SEI layer, respectively. 
The SEI growth decreases the cell capacity directly through the loss of cyclable lithium (also called `LLI' for lost lithium inventory). The formed SEI layer also increases the cell ohmic resistance $R_\text{film}$.

\subsubsection{Electrode particle cracking}
Cell electrodes expand during the intercalation of Li-ions and contract during deintercalation.
This results in alternating stresses, causing initiation and propagation of cracks and loss of active material in the electrodes. Loss of active material results in entrapment and isolation of material containing otherwise useful lithium, leading to capacity fade.
This loss will cause a reduction in capacity due to the entrapment of lithium in isolated sections of the particle and an increase in cell resistance due to a larger overpotential of the positive electrode $\eta^+$.
The capacity loss and crack growth can be modeled using material fatigue models\ \cite{laresgoiti2015modeling}.
We have proposed an advanced model that considers fatigue on a per-cycle basis and includes concentration-dependent stresses\ \cite{pannala_movahedi_garrick_siegel_stefanopoulou_2023}. However, since we are operating within a close range of the tuned conditions, we opted for a simplified version of this model.

For each electrode, we tune the constant fatigue model coefficients $\beta_\text{crack}^\pm$ and exponent $m_\text{crack}$. As shown in Fig.\ \ref{fig: Deg_mechanisms_eq}, the rate of change in active material ratio is calculated by computing the hydrostatic stress at the surface of the particle ($\sigma_\text{h, surf}$). The overall loss is calculated by taking the integral of this rate over battery life.

\subsubsection{Transition-metal dissolution in cathode}

At high cell voltages, particularly near full charge, when the concentration of Li-ions in the cathode is very low, transition metal ions (mostly manganese in NMC cells) from the cathode dissolve into the electrolyte.
Here, we model dissolution by reducing the active material ratio in the cathode $(\varepsilon_{s}^+)$\ \cite{Kindermann2017}, as shown in Fig.\ \ref{fig: Deg_mechanisms_eq}.
Similar to the particle cracking mechanism, loss of active material increases the intercalation overpotential and causes higher resistance.

The dissolution rate varies depending on the chemical composition of the cathode and temperature and can be adjusted by selecting the dissolution exchange current density $i_\text{0,diss}$.  Here, we consider the cathode dissolution mechanism only if we observe a considerable loss of cathode capacity during the calendar aging tests in a cell. The dissolution equilibrium potential in these cells is assumed to be $E_\text{Eq,diss}=4 \ \text{V}$\ \cite{Kindermann2017}.

\subsubsection{Lithium plating}
Lithium plating is another degradation mechanism that can affect the battery lifetime, especially during fast-charging and at lower ambient temperatures. Lithium metal deposits on the surface of the electrode instead of intercalating into it. Similar to the SEI mechanism, plating increases the LLI and ohmic resistance.
Here, we use a modified model for Li-plating that takes into account the non-uniformity of concentration distribution in the electrolyte that is ignored by the SPM\ \cite{pannala_movahedi_garrick_siegel_stefanopoulou_2023}. The only tuning parameter in our Li-plating model is the kinetic rate constant, $k_\text{0,pl}$, as shown in Fig.\ \ref{fig: Deg_mechanisms_eq}.

\subsection{Cells with different dominant degradation mechanisms}
Three different sets of pouch cells that were manufactured and underwent calendar and cycle aging tests at the University of Michigan Battery Lab (UMBL) are considered in this paper. A summary of the characteristics of these cells is in Table \ref{characteristics table}.
\begin{itemize}
\item NMC111\ \cite{Mohtat2021}: These cells have a nominal capacity of 5 Ah. The anode is made of Mag-E3 graphite with a 5\% binder. The cathode consists of NMC111, 3\% carbon black, and 3\% PVDF. We have seen previously that these cells predominantly degrade due to LAM in the negative electrode\ \cite{weng2023differential}, possibly due to the manufacturing process.  
\item NMC622-25C: These cells have a nominal capacity of 3.5 Ah. The anode is made of Superior SLC 1520T Graphite with 1.5\% binder SBR and
1.5\% binder CMC. The cathode consists of single crystal NMC622 with 3\% conductive additive Super C65 and 3\% binder PVDF.
\item NMC622-45C\ \cite{weng2023phenomenological}: These cells are made of the same material as NMC622-25C. They have a nominal capacity of 2.5 Ah, were aged at a higher temperature of 45°C, and underwent a fast formation protocol. These cells represent a case with an extreme SEI growth dominance.

\end{itemize}

The electrolyte for all the cells consisted of ethylene carbonate (EC) and ethyl methyl carbonate (EMC) in a 3:7 weight ratio, $1 mol/lit$ LiPF6, and 2\% vinylene carbonate (VC) additive.

\subsubsection{Description of experimental tests for each digital-twin}

The experimental data for model parameterization includes calendar and cycling aging tests conducted in the University of Michigan Battery Control Lab. Table \ref{test table} presents the aging tests used in this paper.

\begin{table}[htbp]
  \centering
  \footnotesize
    \begin{tabular}{ccccccccc}
    \toprule
          & \multicolumn{4}{c}{Cycle aging} &       & \multicolumn{3}{c}{Calendar aging} \\
\cmidrule{2-9}          & Charge & Disch. & Temp.  & DOD   &   &    & SOC   & Temp.  \\
\cmidrule{2-9} 
    \multirow{2}[1]{*}{NMC111} & C/5   & C/5   & 25 °C & 50\%  &   &    & 1     & 45 °C \\
          & 1.5C  & US06  & 25 °C & 50\%  &   &    & 1     & \ -5 °C \\
    \midrule
    \multirow{2}[1]{*}{NMC622-25} & C/2; 1.3C every 4th cycle& C/2   & 25 °C & 50\% &    &   & 1     & 25 °C \\
          & C/2   & US06  & 25 °C & 100\% &    &   & 0.5   & 25 °C \\
   \midrule          
    NMC622-45   & 1C    & 1C    & 45 °C & 100\% &     &  & 0.9   & 45 °C \\
    \bottomrule
    \end{tabular}%
      \caption{Experimental data for cycle aging and calendar aging tests.}

  \label{test table}%
\end{table}%

Reference performance tests (RPT) were conducted periodically every few weeks for each test condition. The State of Health (SOH) of the cells was estimated using a voltage fitting procedure\ \cite{lee2020electrode}, based on the C/20 data from the RPT tests. The resistance was calculated using hybrid pulse power characterization (HPPC) tests.

\subsection{Fitting Procedure}

\begin{figure}[ht]
    \centering    \includegraphics[width=0.99\textwidth]{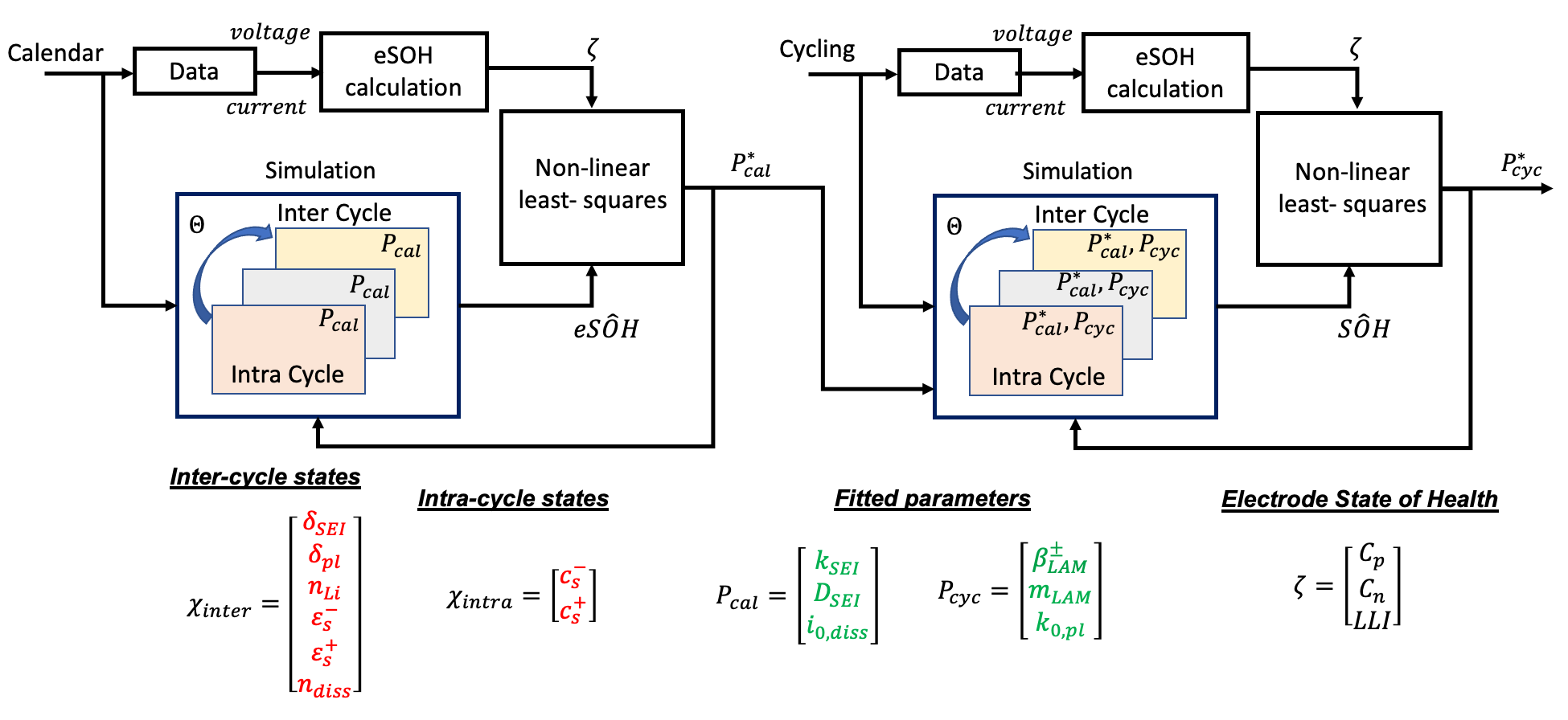}

    \caption{Summary of fitting process. First, the parameters of SEI and cathode dissolution models are found. These parameters, along with cycle test data, are used to find the parameters of Li-plating and mechanical degradation models. An adaptive simulation is used to reduce the simulation time needed for every lifetime iteration to be performed.}
    \label{fig: fitting process}
\end{figure}

Tuning the degradation models based on experimental data similar to the intended cycling conditions is essential for accurate cell degradation prediction. An error minimization process was used to obtain the six parameters of the four degradation mechanisms indicated in green in Fig.~\ref{fig: Deg_mechanisms_eq}. The iterations needed for the minimization to converge, involve multiple lifetime simulations, which is computationally demanding. Although this computation is happening offline in this paper, meaning that laboratory tests and the digital-twin parameterization are not time-constraining, the computational cost is still high. To manage the computational requirements needed for the parameter fitting and for the various simulations, we employed the adaptive accelerated simulation procedure developed in Sulzer et al.\ \cite{sulzer2021accelerated} and used for initial parameterizations by Pannala et al.\ \cite{Pannala2022}. The outline of the fitting procedure is sketched in Fig.~\ref{fig: fitting process}.

\begin{figure}[!ht]
    \centering
    \includegraphics[width=0.99\textwidth]{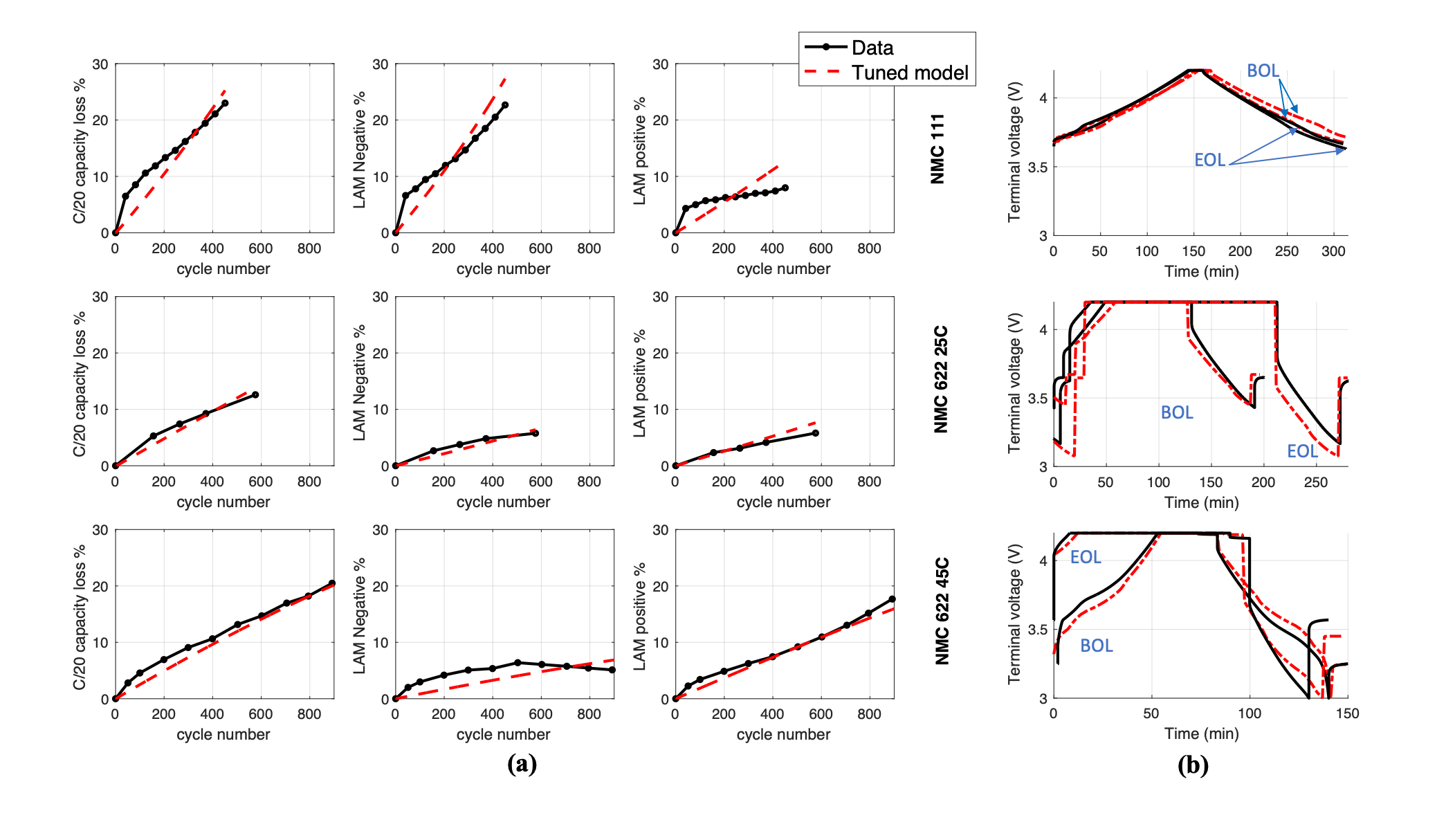}
    \caption{Comparison of the experimental data and fitted model simulations for the three cells. In (a) capacity loss, LAM in the negative and positive electrodes, and (b) voltage behavior of the fresh and aged cells during cycling are shown. For NMC111, the voltage shown is C/5 charge, C/5 discharge at 50\% DOD. For NMC622-25C, the cell is charging and discharging at C/2 at 50\% DOD. NMC622-45C has cycled at 1C charge and discharge rate at 100\%.  
    }
    \label{fig: model & dat}
\end{figure}

For each set of cells, the fitting was performed in consecutive steps.
First, we focus on the parameters governing calendar aging mechanisms.
Calendar aging test results were used to optimize the SEI and dissolution parameters to fit the estimated electrode-specific state of health (eSOH) $\zeta=\begin{bmatrix}
C_\text{p},  
C_\text{n}, 
{LLI} 
\end{bmatrix}^T$, which is associated with decreases in individual electrode capacity $C_\text{p}$, $C_\text{n}$,and the total LLI as shown also in Fig.~\ref{fig: model & dat}.
A derivative-free solver for nonlinear least-squares (DFO-LS)\ \cite{cartis2019improving} and PyBaMM (Python Battery Mathematical Modeling) library\ \cite{sulzer2021python} were used to solve the following optimization problem
\begin{equation}
    \min_{P} \sum_{i=1}^{n_1} \sum_{j=1}^{n_2} w((\zeta_{i,j}-\hat{\zeta}_{i,j})^2,
    \label{eq: optimization}
\end{equation}
where parameters 
\begin{equation*}
P= P_\text{cal} = \begin{bmatrix}
    k_\text{SEI},
    D_\text{SEI},
    i_\text{0,diss}
\end{bmatrix}^T,
\end{equation*}
and $n_1$ is number of conditions tested for each cell, $n_2$ is number of RPTs performed in each test, and $w \in \mathds{R}^3 >0$ is a weight vector\ \cite{movahedi2023physics}. Here, $w=[1, 1, 0.25]$ was chosen to make the parameters in $\zeta$ have the same range.

In the next step, we considered the cycling aging tests and fitted the parameters for Li-plating and mechanical degradation,
\begin{equation*}
P= P_\text{cyc}= \begin{bmatrix}
    k_\text{0,pl},
    \beta_\text{crack}^{+},
    \beta_\text{crack}^{-},
    m_\text{crack}
\end{bmatrix}^T,
\end{equation*}
using the optimization problem \ref{eq: optimization}, and fixing the parameters found in the previous step.

To accelerate the optimization, an adaptive inter-cycle extrapolation technique was used to reduce the required calculations at each iteration\ \cite{sulzer2021accelerated}. 
In this approach, we select specific representative cycles instead of simulating aging for every single cycle. We then extrapolate the states of the degradation models to find the subsequent representative cycles. 
Simulations at every iteration were performed to verify the optimal results of the accelerated problem\ \cite{Pannala2022}. The fitted parameters are presented in Table \ref{tab: fitted parameters}.

\begin{table}[htbp]
  \centering
  \footnotesize

    \begin{tabular}{lccc}
    \toprule
          & \multicolumn{1}{l}{NMC111} & \multicolumn{1}{l}{NMC622-25C} & \multicolumn{1}{l}{NMC622-45C} \\
          \hline
    $k_\text{SEI} (m \ s^{-1})$  & 1.08e-16 & 2.76e-16 & 4.35e-16 \\ 
    $D_\text{SEI} (m^2 \ s^{-1})$ & \multicolumn{1}{l}{1.5.9e-19} & 1.75e-19 & 3.69e-19 \\
    $i_\text{0,diss} (A\ m^{-2})$ & 0     & 0     & 6.24e-04 \\
              \hline
          &       &       &  \\

    $k_\text{0,pl} (m \ s^{-1})$ & 7.42e-10 & 5.48e-10 & 4.64e-10 \\
    $\beta_\text{crack}^+$ & 2.23e-07 & 3.43e-07 & 0.76e-07 \\
    $\beta_\text{crack}^-$& 10.77e-07 & 1.47e-07 & 0.87e-07 \\
    $m_\text{crack}$ & 1.02  & 1.02 & 1.02 \\
    \bottomrule
    \end{tabular}%
  \caption{Fitted parameters for SEI, cathode dissolution, Li-plating, and mechanical degradation models.}
  \label{tab: fitted parameters}%
\end{table}%

The resulting model predictions and experimental data are illustrated in Fig.\ \ref{fig: model & dat}. A close agreement between the eSOH of the developed model and test data can be seen in Fig.\ \ref{fig: model & dat}(a). The comparison between the voltage response of the model and data is presented in Fig.\ \ref{fig: model & dat}(b). With these parameterizations, we simulate various V2G and noV2G scenarios.

\section{V2G and noV2G lifetime simulations} \label{section V2G simulation}

We employed the PyBaMM library to simulate V2G and noV2G duty cycles for each parameterized family of cells.
It is worth noting that without the accelerated simulation used in the tuning section, the full lifetime (800 days) simulation requires approximately 1 hour to complete. However, had we opted for accelerated simulations, the time needed would have been reduced to less than 4 minutes.

\subsection{Simulation results}

\begin{figure}[ht]
    \centering
    \includegraphics[width=0.95\textwidth]{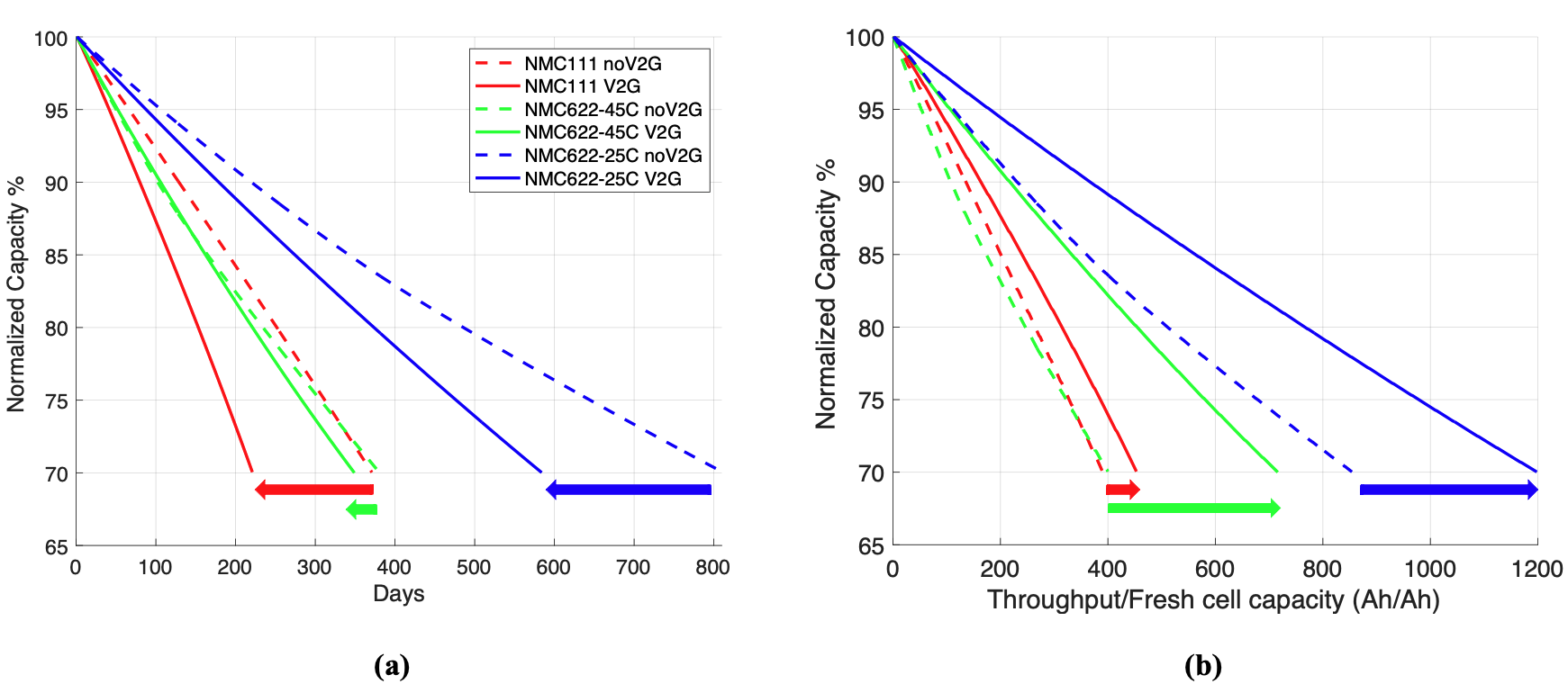}

    \caption{Capacity retention with regard to (a) days and (b) normalized Ah throughput for different scenarios and operational conditions. Discharging the battery to the grid significantly reduces its lifespan in days but only slightly increases Ah throughput in NMC111 cells. On the other hand, V2G in the NMC622-45 case increases Ah throughput with minimal effect on the battery's lifespan in days. The NMC622-25C cell is somewhere between these two cases.}
    \label{fig: capacity retention}
\end{figure}

 The resulting capacity retention in each case is presented in Fig.\ \ref{fig: capacity retention} for days of simulated operation and versus normalized Ah throughput. 
Since the nominal capacities of the considered cells are different, the normalized throughput is defined as:
\begin{equation}
    {\text{Normalized}\; Ah\; \mathrm{throughput}= \frac{Ah\; \text{throughput} }{\text{Nominal capacity}} }
    \label{Normalized thruput}
\end{equation}

As shown in Fig.\ \ref{fig: capacity retention},
the degradation profiles for the NMC111 (cycle aging dominant) cell and NMC622-45C (calendar aging dominant) cell are nearly identical for the noV2G duty cycle. When put through the V2G duty cycle, all the cells experience reduced lifespan and provide additional output. However, based solely on the degradation profile of the noV2G case, one would anticipate similar performance between the NMC111 and NMC622-45C cells under V2G, but they exhibit significant differences. The calendar aging dominant cell delivers much higher throughput and experiences only minimal reduction in longevity. This highlights the importance of detailed cell modeling, which we will discuss further in the next section.

It is also observed that regardless of the cell conditions and chemistries, the degradation order is reversed when compared with days and normalized throughput. In simple terms, the faster the V2G duty cycle causes the cells to age, the less extra throughput it will yield. This trend was also observed by Dubarry et al.\ in their reported experimental results\ \cite{Dubarry2017}.

The summary of the state of each cell at EOL, which is assumed to be the time when each cell reaches 70\% of its initial capacity, is presented in Table \ref{EOL mechanism comparison}. In this table, alongside the eSOH of each cell at the EOL, the contribution of each degradation mechanism to capacity fade is shown.

\begin{table}[h!]
  \centering
  \footnotesize

    \begin{tabular}{p{10em}cccccc}
    \toprule
    \multicolumn{1}{c}{} & \multicolumn{2}{c}{\textcolor[rgb]{ 1,  0,  0}{\textbf{NMC111}}} & \multicolumn{2}{c}{\textcolor[rgb]{ .078,  0,  1}{\textbf{NMC622-25C}}} & \multicolumn{2}{c}{\textcolor[rgb]{ 0,  .69,  .314}{\textbf{NMC622-45C}}} \\
    \midrule
     At 70\% cap. reten.& \multicolumn{1}{c}{\textcolor[rgb]{ 1,  0,  0}{\textbf{noV2G}}} & \multicolumn{1}{c}{\textcolor[rgb]{ 1,  0,  0}{\textbf{V2G}}} & \multicolumn{1}{c}{\textcolor[rgb]{ .078,  0,  1}{\textbf{noV2G}}} & \multicolumn{1}{c}{\textcolor[rgb]{ .078,  0,  1}{\textbf{V2G}}} & \multicolumn{1}{c}{\textcolor[rgb]{ 0,  .69,  .314}{\textbf{noV2G}}} & \multicolumn{1}{c}{\textcolor[rgb]{ 0,  .69,  .314}{\textbf{V2G}}} \\
    \midrule
    Days  & \textcolor[rgb]{ 1,  0,  0}{371} & \textcolor[rgb]{ 1,  0,  0}{221} & \textcolor[rgb]{ .078,  0,  1}{812} & \textcolor[rgb]{ .078,  0,  1}{584} & \textcolor[rgb]{ 0,  .69,  .314}{380} & \textcolor[rgb]{ 0,  .69,  .314}{349} \\
    \midrule
    Norm. thru. (Ah/Ah)& \textcolor[rgb]{ 1,  0,  0}{390.3} & \textcolor[rgb]{ 1,  0,  0}{453.5} & \textcolor[rgb]{ .078,  0,  1}{854.2} & \textcolor[rgb]{ .078,  0,  1}{1198.3} & \textcolor[rgb]{ 0,  .69,  .314}{399.7} & \textcolor[rgb]{ 0,  .69,  .314}{716.1} \\
    \midrule
    \midrule

    $C_\text{p}$ retention \% & \textcolor[rgb]{ 1,  0,  0}{87.5} & \textcolor[rgb]{ 1,  0,  0}{87.0} & \textcolor[rgb]{ .078,  0,  1}{88.1} & \textcolor[rgb]{ .078,  0,  1}{84.1} & \textcolor[rgb]{ 0,  .69,  .314}{77.2} & \textcolor[rgb]{ 0,  .69,  .314}{75.0} \\
    \midrule
    $C_\text{n}$ retention  \% & \textcolor[rgb]{ 1,  0,  0}{75.0} & \textcolor[rgb]{ 1,  0,  0}{68.3} & \textcolor[rgb]{ .078,  0,  1}{91.0} & \textcolor[rgb]{ .078,  0,  1}{86.3} & \textcolor[rgb]{ 0,  .69,  .314}{98.4} & \textcolor[rgb]{ 0,  .69,  .314}{97.0} \\
    \midrule
    LLI\% & \textcolor[rgb]{ 1,  0,  0}{29.5} & \textcolor[rgb]{ 1,  0,  0}{30.0} & \textcolor[rgb]{ .078,  0,  1}{30.0} & \textcolor[rgb]{ .078,  0,  1}{30.0} & \textcolor[rgb]{ 0,  .69,  .314}{29.9} & \textcolor[rgb]{ 0,  .69,  .314}{29.8} \\
    \midrule
    \midrule

    $\text{LLI}_\text{SEI}$\% & \textcolor[rgb]{ 1,  0,  0}{6.5} & \textcolor[rgb]{ 1,  0,  0}{3.3} & \textcolor[rgb]{ .078,  0,  1}{19.1} & \textcolor[rgb]{ .078,  0,  1}{14.1} & \textcolor[rgb]{ 0,  .69,  .314}{23.2} & \textcolor[rgb]{ 0,  .69,  .314}{22.0} \\
    \midrule
    $\text{LLI}_\text{SEI+diss}$\% & \textcolor[rgb]{ 1,  0,  0}{6.5} & \textcolor[rgb]{ 1,  0,  0}{3.3} & \textcolor[rgb]{ .078,  0,  1}{19.1} & \textcolor[rgb]{ .078,  0,  1}{14.1} & \textcolor[rgb]{ 0,  .69,  .314}{27.6} & \textcolor[rgb]{ 0,  .69,  .314}{25.5} \\
    \midrule
   $\text{LLI}_\text{plating}$\% & \textcolor[rgb]{ 1,  0,  0}{1.4} & \textcolor[rgb]{ 1,  0,  0}{1.5} & \textcolor[rgb]{ .078,  0,  1}{2.9} & \textcolor[rgb]{ .078,  0,  1}{3.5} & \textcolor[rgb]{ 0,  .69,  .314}{0.6} & \textcolor[rgb]{ 0,  .69,  .314}{0.9} \\

    \midrule
    $\text{LLI}_\text{crack}$\% & \textcolor[rgb]{ 1,  0,  0}{21.6} & \textcolor[rgb]{ 1,  0,  0}{25.3} & \textcolor[rgb]{ .078,  0,  1}{8.3} & \textcolor[rgb]{ .078,  0,  1}{12.7} & \textcolor[rgb]{ 0,  .69,  .314}{1.7} & \textcolor[rgb]{ 0,  .69,  .314}{3.5} \\
        \midrule
    $\text{LLI}_\text{LAM}$\% & \textcolor[rgb]{ 1,  0,  0}{21.6} & \textcolor[rgb]{ 1,  0,  0}{25.3} & \textcolor[rgb]{ .078,  0,  1}{8.3} & \textcolor[rgb]{ .078,  0,  1}{12.7} & \textcolor[rgb]{ 0,  .69,  .314}{6.1} & \textcolor[rgb]{ 0,  .69,  .314}{6.9} \\
    \bottomrule
    \end{tabular}%
    \caption{State of health and contributions of different degradation mechanisms at the EOL (70\% capacity retention) for V2G and noV2G scenarios. Compared to the noV2G case, the mechanical degradation increases, and SEI contribution decreases in the presence of V2G services. The SEI growth and cathode dissolution constitute calendar aging degradation. LLI due to LAM is the summation LLI in electrodes either because of mechanical degradation or cathode dissolution. }
    
  \label{EOL mechanism comparison}%
\end{table}%

\subsection{Dominant degradation mechanisms}

To compare the relative contribution of each degradation mechanism in each case, we can calculate the share of each mechanism as a fraction of the total loss of lithium inventory (LLI).
This is a reasonable assumption since the cells do not experience a knee in their degradation profile, and the electrode capacities do not become a limiting factor.
The breakdown of these fractional contributions, along with the state of health at the EOL, is presented in Table \ref{EOL mechanism comparison}. The total LLI at each cycle (day) is calculated as:
\begin{equation}
    {\text{LLI}_k=\frac{\Delta n_\text{Li}}{n_\text{Li,BOL}}= \frac{n_\text{Li,BOL}-n_\text{Li,k}}{n_\text{Li,BOL}} }
    \label{LLI def}
\end{equation}
where $n_\text{Li,BOL}$ and $n_\text{Li,k}$ are the moles of cyclable lithium at the beginning of life (BOL) and at $\text{cycle}=k$ of the cells. Here, we assume that the battery will last until the capacity reaches 70\% 

\begin{equation}
    {\text{LLI}= \text{LLI}_\text{SEI}+\text{LLI}_\text{plating}+\text{LLI}_\text{diss}+\text{LLI}_\text{crack}}
    \label{LLI divided}
\end{equation}

In most situations, LAM can reduce the capacity of a battery due to LLI ($\text{LLI}_\text{LAM}$). The LAM, due to particle cracking and cathode dissolution, creates electrically isolated particles that can no longer be charged or discharged. Since the lithium ions cannot be released without allowing an electron to reach the current collector, the lithium becomes trapped at the time of the failure, reducing the battery's usable lithium supply and thereby reducing its overall capacity. It has been shown\ \cite{sulzer2021accelerated} that the amount of LLI due to LAM can be calculated using the following formula:

\begin{equation}
    {\frac{dn_\text{Li, LAM}}{dt} = \frac{3600}{F}\left(x \frac{dC_{n}}{dt}+y\frac{{dC}_{p}}{dt}\right)}
    \label{eq: LLI_LAM}
\end{equation}
where $x$ and $y$ are the stoichiometry of the negative and positive electrodes, respectively.
The LLI due to LAM is the summation of LLI due to cathode dissolution and particle cracking $(\text{LLI}_\text{LAM}=\text{LLI}_\text{diss}+\text{LLI}_\text{crack})$. The cathode dissolution and particle cracking portions of the degradation are calculated as:

\begin{equation}
\label{eq: LLI diss}
    {\text{LLI}_\text{diss}=\frac{  l^+ \,c_{s,max}^+ \, \int_\text{BOL}^\text{EOL} y\frac{d\varepsilon_\text{diss}^+}{dt} \,dt }
    {\Delta n_\text{Li}}} 
\end{equation}
\begin{equation}
\label{eq: LLI crack}
    {\text{LLI}_\text{crack}=\frac{  l^+ \,c_{s,max}^+ \, \int_\text{BOL}^\text{EOL} y\frac{d\varepsilon_\text{crack}^+}{dt} \,dt  + l^- \,c_{s,max}^- \, \int_\text{BOL}^\text{EOL} x\frac{d\varepsilon_\text{crack}^-}{dt} \,dt}
    {\Delta n_\text{Li}}}
\end{equation}

where $\varepsilon_\text{diss},\  \varepsilon_\text{crack},  \ l,$ and $c_{s,max}$ denote the change in active material ratio due to dissolution and particle cracking, electrode thickness, and maximum concentration in each electrode, respectively.
 As seen from this formula, the amount of trapped lithium is proportional to the rate of LAM ($\dot{C_p} $ and $\dot{C_n}$) and the amount of lithiation in each electrode when LAM occurs.

The SEI and plating contributions are calculated as follows:
\begin{align}
\label{eq: LLI SEI}
    {\text{LLI}_\text{SEI}= \frac{ \int_\text{BOL}^\text{EOL} j_\text{SEI} \,dt /F}{\Delta n_\text{Li}}} \\ 
    \label{eq: LLI plating}
    {\text{LLI}_\text{plating}= \frac{ \int_\text{BOL}^\text{EOL} j_\text{pl} \,dt /F}{\Delta n_\text{Li}}}
\end{align}
where $j_\text{SEI}$ and $j_\text{pl}$ are the SEI and Li-plating current densities respectively.

We define the LLI due to calendar aging as the summation of LLI due to SEI and cathode dissolution $(\text{LLI}_\text{Cal}=\text{LLI}_\text{SEI}+\text{LLI}_\text{diss})$.

Analyzing the resulting breakdown of contributions from each degradation mechanism in Table \ref{EOL mechanism comparison}, it can be seen that in all three cases, performing V2G load-shifting services elevates the relative share of mechanical degradation and Li-plating to the total capacity fade. This observation is logical, considering that these degradation mechanisms intensify with higher Ah throughput by the EOL. 

Performing V2G, however, decreases the relative contribution of calendar aging mechanisms (SEI growth and cathode dissolution). The reason behind this reduction is two-fold. First, the degradation attributed to the other two mechanisms has intensified. Second, the cell degrades faster in time, leading to an earlier EOL in time. Since SEI growth and cathode dissolution mechanisms are known to be increasing functions of time\ \cite{Kindermann2017}, the amount of LLI due to these mechanisms will be lower. Now, we review the degradation mechanisms in each individual cell family.

\subsubsection{NMC111: Dominant $\text{LAM}_\text{Neg}$ degradation}
Analyzing Table \ref{EOL mechanism comparison} shows that the dominant degradation mode for NMC111 cells is LAM caused by particle cracking. For the noV2G scenario, 73\% (21.6/29.5) of the LLI is the result of mechanical degradation. This portion for the V2G case rises to 86\% (25.3/30.0) as shown in the $\text{LLI}_\text{crack}$ row of the Table. 
The capacity retention of the electrodes ($C_\text{p}$ and $C_\text{n}$) at EOL
shows that most of this LLI is due to particle cracking in the negative electrode. 

\subsubsection{NMC622-45C: Mostly SEI degradation}
Due to being tuned at \SI{45}{\celsius} for both calendar aging and cycling conditions, these cells suffer from degradation predominantly caused by SEI growth.
In the noV2G case, 78\% (23.2/29.9) and in the  V2G case, 74\% (22.0/29.8) of the capacity fade corresponds to SEI ($\text{LLI}_\text{SEI}$ row in Table \ref{EOL mechanism comparison}).  

Due to the cathode dissolution in this cell, $\text{LAM}_\text{Pos}$ is considerable, as shown in the $C_\text{p}$ retention row in Table \ref{EOL mechanism comparison}. However, this $C_\text{p}$ fade has not translated to large values of LLI due to LAM ($\text{LLI}_\text{LAM}$ row in Table \ref{EOL mechanism comparison}).
This contradictory observation becomes clear when we examine the rate of change in the moles of lithium entrapped in the electrodes as the electrode loses its active material from Equation \ref{eq: LLI_LAM}.

Since the dissolution only happens at higher voltages (corresponding to smaller values of $y$ and the positive electrode being almost empty of lithium), $y\frac{dC_\text{p}}{dt}$ is relatively small. In other words, at higher voltages, the amount of Li-ions in the positive electrode is minimal. Therefore, the loss of active material in the positive electrode will not isolate a large amount of lithium.
Overall, these cells have the highest portion of calendar aging (SEI + dissolution) in the capacity fade compared to the other two cell families.

\subsubsection{NMC622-25C: Mixed degradation}
Unlike the previous cell families, in these cells, the SEI growth and mechanical degradation contributions to the capacity fade are comparable.
The lower temperature of cycling and calendar aging conditions in the experimental data used to tune the physics-based model in these cells results in reduced SEI growth compared to NMC622-45C cells. The mechanical degradation in these cells is lower compared to NMC111 cells owing to the improved anode chemistry.
The degradation in these cells can be seen as a mid-point between two extreme cases represented by the NMC111 and NMC622-45C cells.

\section{Analysis of V2G for each cell family}

Now that we have explained the physics-based model and examined the V2G results for cell families with different dominant degradation modes, we discuss the impact of V2G services on the degradation of these cells and specifically shed more light on the prior published results.  
Examining the capacity retention results in Fig.\ \ref{fig: capacity retention}, we can draw the following conclusions.

The cell with NMC111 cathode, which is dominated by $\text{LAM}_\text{Neg}$, degrades at a faster rate when it is subjected to V2G compared to the other cells. This considerable loss of battery life in time does not result in a substantial extra Ah throughput.
This trend was also observed in experimental studies conducted by Dubarry et al.\ \cite{Dubarry2017} and Peterson et al.\ \cite{peterson2010lithium}, which makes it likely that their cells are prone to degradation mechanisms that get amplified due to the additional Ah throughput from V2G, such as particle
cracking as the primary source of degradation. 

The NMC622-45C cell family, dominated by calendar aging, shows only a slightly faster degradation in time when subjected to V2G. 
However, these cells show substantially more throughput (extra virtual mileage) when subjected to V2G.
This demonstrates that a considerable amount of unused mileage in these cells will go to waste if they are not used for V2G. 
The same trend has been reported in studies like Kim et al.\ \cite{Kim2022} and Wei et al.\ \cite{wei2022comprehensive}. The cells in these previous studies were most likely prone to calendar aging mechanisms as well.

The NMC622-25C cells can be considered a mid-point between the other two extreme cases. Specifically, when subjected to V2G services, they have a moderate loss of life increase in time and a mild increase in throughput, which can be related to the comparable fractions of LAM- and SEI-related degradation in these cells. 

These three cases suggest a close relation between the ratio of benefits to damage of V2G and the ratio of the calendar to the total aging. To explore this relation, we introduce the V2G `Throughput gained vs. Days lost' ratio. 

\subsection{Relationship between V2G Throughput gained vs. Days lost (TvD) and the calendar aging in cell lifetime}

It has been known that extra Ah throughput is a major factor in increased battery degradation. 
The relative reduction of life for every Ah increase, as was shown in this paper, is not equal for all the cell families. 
Additionally, we showed that the amount of extra Ah throughput that can be extracted from the cells through V2G services before a particular capacity fade is reached depends on their chemistry and, more specifically, their dominant degradation mechanisms.

To quantify the damage and benefits of V2G services compared to a baseline case (without V2G) for each cell condition, we define the following metric:

\begin{equation}
\mathrm{V2G\; TvD\;  ratio} =\frac{\mathrm{Life\; gained\; in\; use\;} Ah/Ah\; \%}{ \mathrm{Life\; lost\; in\; days\;} \%} 
    \label{benefit ratio}
\end{equation}
In this metric, we use relative life lost in the denominator (i.e., days lost divided by baseline noV2G life in days) to reflect the harm to the battery. The extra relative Ah throughput in the numerator (i.e., improved throughput divided by baseline noV2G throughput) represents the gain or benefit. To make this definition comprehensive, in exceptional cases where V2G fails to yield any additional throughput due to an extreme loss of life (numerator smaller than zero), we define TvD=0. If V2G increases battery life (negative denominator), we presume TvD$\rightarrow \infty$.

It is worth noting that instead of using Ah throughput, one can use Wh throughput in this definition, which might be more useful for certain applications. Simulations indicated that for the cases considered here, Ah and Wh are equivalent, and the difference in TvD values is minimal.

The TvD metric will have a large value when the additional extracted Ah throughput due to V2G operations is large, but the battery life in days is not reduced substantially by V2G. In other words, the battery might be over-engineered for the original use case.
On the other hand, if the V2G services cause a significant loss in the life of the battery in days, but only give a small amount of extra throughput, the TvD ratio will be small.

\begin{figure}[h]
    \centering
    \includegraphics[width=0.99\textwidth]{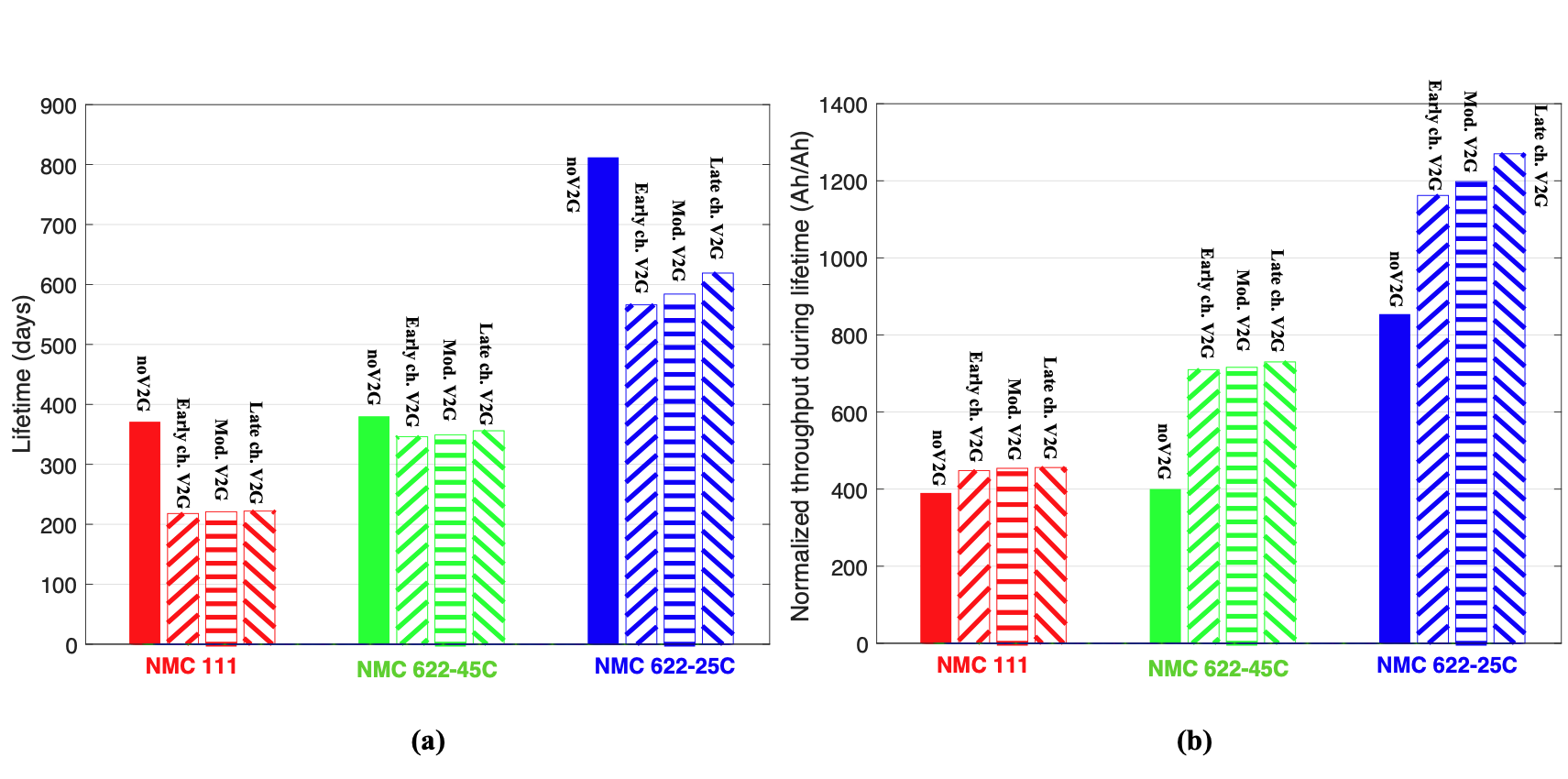}
    \caption{
The lifetime of three cell families in (a) days and (b) integrated current in Ah throughput when subjected to regular driving duty cycle (noV2G) and moderate V2G. The lifetime in days and Ah throughput are also shown for the cases where the charging in the V2G cases happens immediately after discharge to the grid (called early-charge V2G) and where the charging happens right before driving (called late-charge V2G) duty cycles.
    }
    \label{fig:barcharts}
\end{figure}

\begin{figure}[h]
    \centering
    \includegraphics[width=0.99\textwidth]{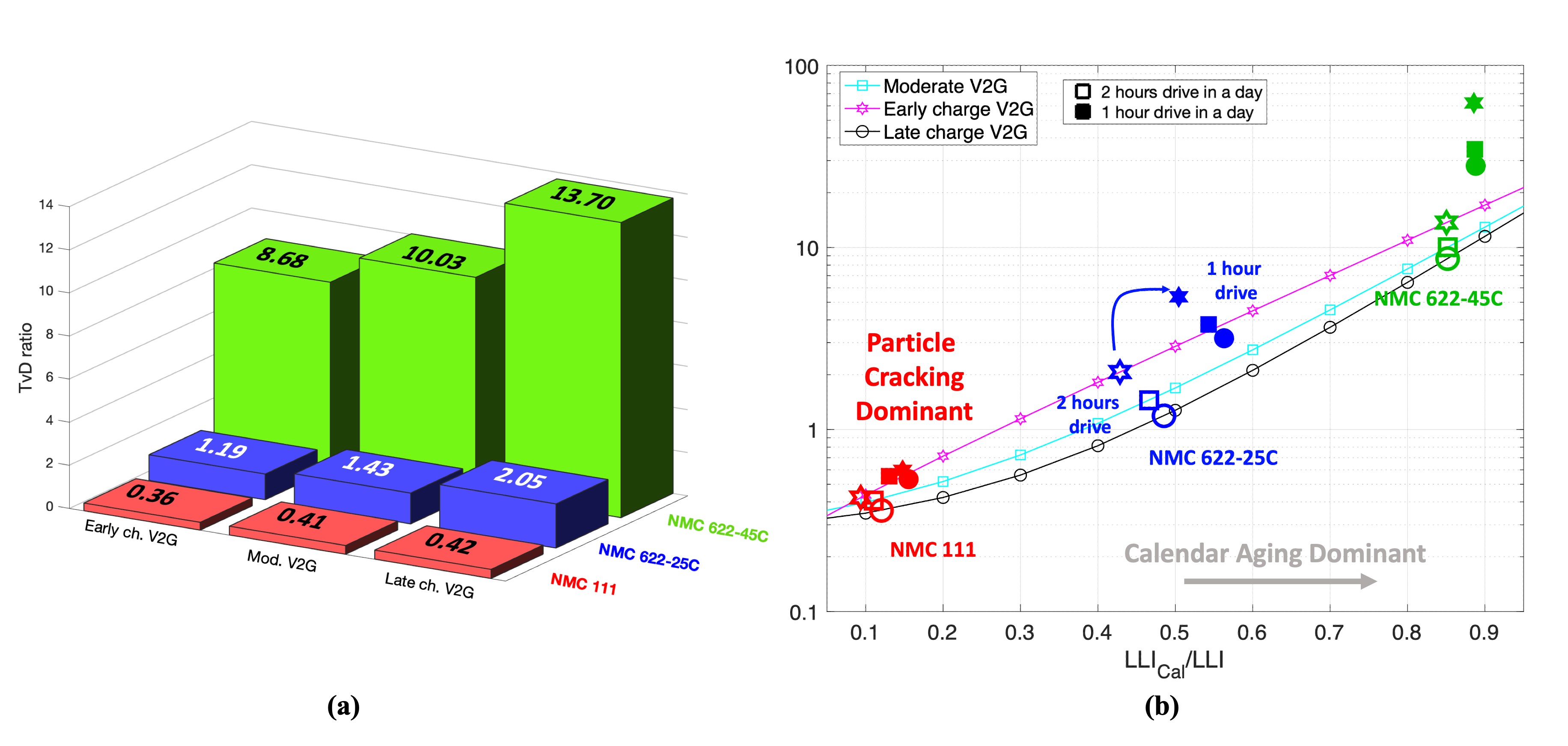}
    \caption{
    (a) TvD ratio for different V2G operations for different cell families. (b) Comparison of the V2G TvD ratio in logarithmic scale for moderate, late-charge, and early-charge V2G based on the portion of LLI that is caused by calendar aging (Red points represent NMC111 cells, blue points are NMC622-25C, and green points represent NMC622-45C cells.).
    The effect of driving distance on the TvD ratio is also shown (filled markers show 2 hours/day, and empty markers show 1 hour/day driving). 
    }
    \label{fig: 3d bar charts and curve}
\end{figure}

The resulting EOL battery age and Ah throughput for noV2G and V2G cases (presented as moderate V2G) are presented in Fig.\ \ref{fig:barcharts}  for the three cell families. The TvD ratio is shown in Fig.\ \ref{fig: 3d bar charts and curve}(a).
As can be seen, NMC622-45C cells have the highest TvD because the SEI will degrade them even during parking, so we might as well use them for V2G. They are followed by NMC622-25C cells. The NMC111 cells have the lowest TvD ratio since Ah throughput kills these cells.

The SEI growth and cathode dissolution aging effects are known to be functions of time\ \cite{safari2008multimodal, Kindermann2017}, so regardless of whether the cell is used or not,  SEI and dissolution will continue to degrade the cell. 
On the other hand, degradation mechanisms like particle cracking and lithium plating only occur when the cell is cycling.
Hence, when the battery is degrading primarily due to calendar aging mechanisms (like NMC622-45C cells here), V2G will be beneficial for extracting the unused potential of the battery. On the other hand, when the cell is mainly degrading due to cycling mechanisms such as particle cracking (similar to NMC111 cells here), most of the possible Ah throughput is already being used by driving the vehicle, and adding V2G services will just degrade the cell faster. Therefore, the V2G TvD ratio for these cells is smaller. 

To determine whether a discernible pattern exists for car owners regarding the viability of V2G  services based on the usage of the battery, we analyze the TvD metric in Fig.\ \ref{fig: 3d bar charts and curve}(b).
This analysis focuses on the impact of calendar aging ($\text{LLI}_\text{SEI+diss}$ row in Table \ref{EOL mechanism comparison}) on overall degradation $(\text{LLI}_\text{Cal}/\text{LLI})$. 
This figure clarifies the significance of calendar aging $(\text{LLI}_\text{Cal})$ portion of the overall capacity fade (LLI) on the TvD ratio.

The TvD ratio has a semilog relationship with $\text{LLI}_\text{Cal}/\text{LLI}$. It can be seen that as the calendar aging component of the degradation increases, V2G services become more advantageous per the high TvD. This means that the benefit in throughput outweighs the loss in days, and V2G can harness the untapped value of the battery.

Considering these findings, EV manufacturers can determine how additional V2G throughput should impact their warranty policies by determining what mechanisms are aging their batteries. If their battery technology is mostly affected by cycling aging, they should include the additional throughput as virtual miles within the warranty limit, possibly with a higher weight, to discourage V2G use. Conversely, suppose the battery is predominantly affected by calendar aging. In that case, they should consider ways to encourage participation in V2G services, as it could lower the overall cost of ownership by the V2G compensation without increasing the need for premature battery replacement.

What we have shown here emphasizes the necessity of considering calendar aging and cycle aging of cells separately. 
 The physics-based model, as indicated in this work, delivers more accurate results. Still, in case one decides to accept the lower accuracy and utilize empirical models, it is essential to tune and empirically model the calendar aging and cycle aging separately and take into account the contribution of each mode of degradation to be able to predict degradation and optimize the usage profile of batteries during V2G.

 Beyond the calendar aging component, other important factors for TvD are the intensity of the drive cycle, times parked without charging, operating voltage range, and the type of V2G services. 
 Our physics-based digital-twin enables us to evaluate various scenarios and account for different secondary factors affecting TvD. To demonstrate this capability, we analyzed the impact of the driving distance of the drive cycles and the average SOC of the duty cycles on TvD for three different cell families.

\subsection{Average SOC effect}

To analyze the effect of $\text{SOC}_\text{ave}$ on TvD, similar to the experiments conducted by Kim et al.\ \cite{Kim2022} and Peterson et al.\ \cite{peterson2010lithium}, we use the created model to simulate the degradation in different $\text{SOC}_\text{ave}$ values. 
We show that while $\text{SOC}_\text{ave}$ at which the battery operates is an important factor to consider, it may not have as much impact as other factors, such as the propensity to calendar aging or driving distance on the TvD.

Specifically, we repeated the V2G simulations for two additional scenarios: one where the charging happens right after the driving periods (early-charging) and the other where the charging happens right before driving (late-charging). The description of these new scenarios, along with moderate V2G that was discussed thus far, was presented earlier in Section \ref{V2G service section}. The SOC profiles were presented in Fig.\ \ref{fig: SOCs}. 

In the early-charging case, the cell rests when fully charged and has a higher average SOC ($\text{SOC}_\text{ave}=0.88$) than in the moderate case($\text{SOC}_\text{ave}=0.79$). In the late-charging V2G case, the resting period happens at a low SOC; therefore, the average SOC is lower ($\text{SOC}_\text{ave}=0.61$)than in other scenarios. The resulting TvD ratios for each cell family and different charging times are presented in Fig.\ \ref{fig: 3d bar charts and curve}(a).

During late-charging, TvD has the highest value, and during early-charging, it has the lowest value, but the sensitivity of TvD to charging times varies among cell families.  When it comes to NMC622-25C and NMC622-45C cells, TvD is highly sensitive to the charging protocol's average SOC, while NMC111 cells are the least sensitive to charging time and $\text{SOC}_\text{ave}$.

To further investigate the results, the lifetime and Ah throughput of the cells under different duty cycles are presented in Fig.\ \ref{fig:barcharts}. 
As is expected, late-charging (lower $\text{SOC}_\text{ave}$) reduces the degradation rate, and early-charging (higher $\text{SOC}_\text{ave}$) increases degradation. The amount of this change is, however, different for different cell chemistries and conditions. The main reason behind these changes is the calendar aging mechanisms. Both SEI growth and cathode dissolution increase when the SOC is set at higher levels. Hence, NMC111 cells are not substantially affected by the variation of $\text{SOC}_\text{ave}$ as they are primarily degrading due to mechanical degradation, as discussed in Section \ref{section V2G simulation}.

The effect of charging scheduling on degradation is more noticeable in NMC622-25C and NMC622-45C cells due to their significant SEI. 
Note also that NMC622-45C cells have both SEI and cathode dissolution contributions to $\text{LLI}_{\text{Cal}}$, making the effect of SOC on TvD more complex.

In Fig.\ \ref{fig: 3d bar charts and curve}(b), we present the dependencies of the TvD ratio for different V2G scenarios on the computed ratio of $\text{LLI}_\text{Cal}/\text{LLI}$ during V2G cases. These trends support our previous claim that there is a direct relationship between $\text{LLI}_\text{Cal}$ and the TvD ratio of different load-shifting V2G services for all charging patterns, which ultimately determines their feasibility.

In a similar analysis, we consider the range of SOC windows as an important variable to explore in V2G. 
Suppose we lower both the upper and lower limits of the SOC, according to Equation \ref{eq: LLI_LAM}. In that case, it becomes clear that reducing the SOC window is beneficial for cells where the negative electrode experiences a significantly greater loss of active material ($\text{LAM}_{\text{Neg}}$) than the positive electrode ($\text{LAM}_{\text{Pos}}$). As both SEI and mechanical degradation will diminish with a lower voltage window. For cells with larger LAM in the positive electrode, the extent of this increased advantage may not be particularly evident. 
This shows that our digital-twin can identify variables that impact TvD and be used to optimize it while accounting for diverse EV driver behavior.

\subsection{Driving distance}

To examine how driving distance and time affect TvD, we analyzed a shorter distance driven from home to work and back. This shorter drive cycle includes half of a UDDS drive cycle, a HWFET drive cycle, and half of a US06 drive cycle. Compared to the longer drive cycle, this shorter drive cycle takes only 30 minutes (instead of 1 hour), equivalent to a 17-mile drive (instead of 34.1 miles) per trip. The total distance of the two-way trip of the shorter drive cycle is similar to the average daily distance driven by EVs\ \cite{DONG201444}.
The resulting TvD of the three cell families using the shorter driving distance is shown and analyzed in Fig.\ \ref{fig: 3d bar charts and curve}(b).



As expected, more time spent parked increases the portion of LLI during calendar aging ($\text{LLI}_{\text{Cal}}$), which consequently increases the TvD ratio.
The increase in $\text{LLI}_{\text{Cal}}$ is due to the decrease in diving time, resulting in less throughput, with a secondary influence from a rise in the proportion of calendar aging in the LLI due to higher average SOC. The increase in TvD is more noticeable in cells that are dominated by calendar aging, as using them less leaves more unused capacity in these cells. This trend is visible across all charging patterns as well.

Based on the observed sensitivity of the TvD to the driving distance, we can draw a more general conclusion. Applications that require extensive driving, such as heavy-duty electric vehicles or ride-sharing, will have a smaller TvD and, therefore, minimal opportunities to offer V2G services.

 \section{Conclusion and future work}

In this paper, we utilized an experimentally tuned physics-based models as  digital-twins to investigate how the contribution of various aging mechanisms to capacity fade impacts the cost and benefit of load-shifting V2G services.
To achieve this, three sets of experimental data with varying susceptibilities to degradation modes were examined to tune the mechanisms of particle cracking, SEI growth, lithium plating, and cathode dissolution.
Then, we considered these three different sets of Li-ion battery families in the presence of V2G services. 
Degradation simulations of V2G and baseline scenarios were performed, and contributions of each mechanism on capacity fade for each case were compared.
The advantage of providing V2G services in extra Ah throughput before 70\% capacity retention is reached then was quantified. The V2G Throughput gained vs. Days lost (TvD) ratio was introduced by clarifying the associated normalized reduction in life (in days).

Based on the found TvD values, we conclude that V2G services tend to be more beneficial for cells experiencing light driving loads and having a substantial Ah throughput headroom. This observation suggests that the net positive impact of V2G is pronounced under conditions of lower operational stress and higher capacity margins, particularly when SEI will age the cells significantly while parked.

To this end, we identified a non-dimensional metric ($\text{LLI}_\text{Cal}/\text{LLI}$) that informs car owners about the viability of V2G services based on the cell's fundamental aging characteristics. We observed a nearly linear relationship between the logarithm of TvD and $\text{LLI}_\text{Cal}/\text{LLI}$. Therefore, using this metric, we showed that for the cell chemistries and conditions with a higher calendar aging contribution to capacity degradation, V2G is more beneficial than harmful. This finding can help EV manufacturers decide how V2G services affect their warranty depending on the level of degradation dominance of the batteries.

With the progress in cell technology, especially improvements in the mechanical characteristics of electrodes, batteries that primarily degrade during storage rather than cycling can offer benefits by delivering V2G services. 
This benefit can potentially overshadow the additional aging caused by V2G. 

Important extensions of our work include the estimation of the degradation mechanism contributions to capacity fade such that car and fleet owners can compute their battery's dominant degradation via a real-time $\text{LLI}_\text{Cal}/\text{LLI}$ estimate. Our near-term goal is to parameterize our digital-twin for LFP batteries where $\text{LLI}_\text{Cal}/\text{LLI}$ is expected to be significant. Our next step is parameterizing this digital-twin for cells with high silicone content.

Exploring additional applications for the energy stored in vehicle batteries, such as vehicle-to-building or vehicle-to-load, are tasks the growing community of V2G application researchers can investigate with our open-source digital-twin.

\newpage
\textbf{Statement}: During the preparation of this work, the authors used Grammarly in order to improve the readability. After using these tools, the authors reviewed and edited the content as needed and take full responsibility for the content of the publication.

\section*{Funding} 
This work was supported by the Coordinating Research Council grant SM-E-4/8.

\section*{Data} 

It will be provided along with codes.

\section*{CRediT authorship contribution statement}
\textbf{Hamidreza Movahedi:} Methodology, Software, Investigation, Writing - Original Draft, \textbf{Sravan Pannala}: Software, Writing - Review \& Editing, \textbf{Jason Siegel}: Conceptualization, Investigation,\textbf{Stephen J. Harris}: Conceptualization, \textbf{David Howey}: Conceptualization, Writing - Review \& Editing,  \textbf{Anna Stefanopoulou}:  Conceptualization, Writing - Review \& Editing, Supervision
\newpage
\appendix

\section{Glossary of terms}
\scriptsize
\begin{longtblr}[
    caption={Description of Model Variables and Parameters}
    ] 
    {
  colspec = {XX[4]},
  hlines,
} 
        
        Variable/Par. & Description  \\
        $A$ & Area of electrode [${m}^2$]\\
        $a_{s}$ & Surface area to volume ratio [$m^{-1}$] \\

        $\alpha$ & Charge transfer coefficient\\
        $\alpha_\text{SEI}$ & Charge transfer coefficient of SEI formation \\

        $\beta_\text{crack}$ & Electrode Cracking rate [$s^{-1}$] \\
        $C$ & Cell capacity [Ah] \\
        $C_p , C^+$  & Positive electrode capacity [Ah] \\
        $C_n , C^-$  & Negative electrode capacity [Ah] \\
        
        $c_{e}$ & Conc. of Li in electrolyte [$mol \; m^{-3}$]\\
        $c_{s}$ & Conc. of Li in electrode  [$mol \; m^{-3}$]\\
        $c_\text{SEI}$ & Conc. of SEI layer in electrode  [$mol \; m^{-3}$]\\
        $c^{s}_\text{EC}$ & Conc. of solvent in electrolyte  [$mol \; m^{-3}$]\\
        $c_{s,avg}$ & Average conc. of Li in particle  [$mol \; m^{-3}$]\\
        $c_{s,max}$ & Maximum conc. of Li in electrode  [$mol \; m^{-3}$]\\
        $c_{ss}$ & Conc. of Li at particle surface  [$mol \; m^{-3}$]\\

        $D_{s}$ &  Electrode diffusion coefficient  [$m^2 \; s^{-1}$]\\
        $D_\text{SEI}$ & SEI layer diffusivity [$m^2 \; s^{-1}$]\\
        $\delta_\text{SEI}$ & Thickness of SEI layer [$m$]\\
        $\delta_\text{pl}$ & Thickness of plated lithium [$m$]\\
        
        $E$ & Young's modulus of the electrode material [$Pa$] \\
        $E_\text{Eq,diss}$ & Dissolution equilibrium potential [$V$] \\
        $\varepsilon_{s}$ & Active material ratio\\
        $\varepsilon_\text{diss}$ &  Effect of dissolution on active molar ratio\\
        $\varepsilon_\text{crack}$ & Effect of particle cracking on active molar ratio \\
        
        $\eta$ & Bulter-Volmer overpotential [$V$] \\
        $\eta_\text{diss}$ & Overpotential of cathode dissolution [$V$]\\
        $\eta_\text{pl}$ & Overpotential of plating [$V$]\\
        $\eta_\text{SEI}$ & Overpotential of SEI formation [$V$]\\
        $F$ & Faraday's constant [$C \;mol^{-1}$]\\

        $i_0$ & Reference exchange current density of electrode [$A \; m^{-2}$]\\
        $i_\text{0,diss}$ & Reference exchange current density of dissolution [$A \; m^{-2}$]\\
        $i_\text{0,pl}$ & Reference exchange current density of lithium plating [$A \; m^{-2}$]\\
        
        $j_{int}$ & Intercalation current density  [$A \; m^{-2}$]\\
        $j_\text{pl}$ & Current density of lithium plating [$A \; m^{-2}$]\\
        $j_\text{SEI}$ & Current density of SEI formation [$A \; m^{-2}$]\\
        $j_\text{tot}$ & Total current density in the electrode [$A \ m^{-2}$]\\

        $k_\text{SEI}$ & SEI kinetic rate constant [$m \; s^{-1}$]  \\
        $k_0$ & Exchange current density [$\mathrm{A}\mathrm{m}^{-2}(\mathrm{m}^3\mathrm{mol})^{-1.5}$]\\
        $\kappa_\text{SEI}$ & Ionic conductivity of SEI [${S} \; {m}^{-1} $] \\
        $\kappa_\text{pl}$ & Ionic conductivity of plated lithium [${S} \; {m}^{-1} $] \\
        
        $l$ & Length of electrode [$m$]\\
        
        $M_\text{SEI}$ & Molar conc. of SEI layer [$mol \;{m}^{-3}$]\\  
        $m_\text{crack}$ & LAM exponent \\
        
        $n_\text{Li}$ & Amount of cyclable Li [$mol$] \\
        $\nu$ & Poisson's ratio\\
        
        $\Omega$ & Partial molar volume [${m}^3 \;mol^{-1}$]\\
        $\Omega_\text{SEI}$ & SEI partial molar volume [${m}^3 \;mol^{-1}$]\\
        $\Omega_\text{pl}$ & Plated lithium partial molar volume [${m}^3 \;mol^{-1}$]\\
        
        $\phi_s$ & Potential of electrode [$V$]\\
        
        $\sigma_\text{critical}$ & Critical stress of the electrode [$Pa$]\\
        $\sigma_{h}$ & Hydrostatic stress in the particle [$Pa$]\\
        
        $R$ & Specific gas constant [$J \;mol^{-1}$]\\
        $R_\text{p}$ & Radius of particle [m]\\ 
        $\rho_\text{SEI}$ & Density of SEI layer [$kg \;{m}^{-3}$] \\
        
        $T$ & Temperature of battery [$K$]\\
        
        $U$ & Open circuit potential of electrode [$V$]\\
        $U_\text{SEI}$ & Potential of SEI formation [$V$]\\

        $V$ & Terminal voltage [$V$] \\
        
        $V_{R}$ & Voltage drop across film resistance [$V$]\\

        $x$ & Stoichiometry of the negative electrode \\
        $y$ & Stoichiometry of the positive electrode \\
\end{longtblr}

\newpage

\section{Model parameters} \label{appendix.SPM papameters}

\normalsize
\begin{table}[h!]
  \centering

    \begin{tabular}{lccc}
    \toprule
          & \multicolumn{1}{l}{NMC111} & \multicolumn{1}{l}{NMC622-25C} & \multicolumn{1}{l}{NMC622-45C} \\
          \hline
    $\alpha$ & 0.5 & 0.5 & 0.5\\
    $\alpha_\text{SEI}$ & 0.5 & 0.5 & 0.5\\
    
    $c_{s,max}^{+} [mol\; m^{-3}]$  & 35380 & 33700 & 37500\\
    $c_{s,max}^{-} [mol\; m^{-3}$  & 28746 & 27200 & 28746\\

    $D_{s}^{+} [m^2\; s^{-1}]$  & 8e-15 & 8e-15 & 8e-15\\      
    $D_{s}^{-} [m^2 \;s^{-1}]$  & 8e-14 & 8e-14 & 8e-14\\

    $E_\text{Eq,diss} [V]$& - & - & 4.0\\

    $\kappa_\text{SEI} ([\Omega m]$& 3e4 & 1.3e3 & 3e4\\

    $\Omega_\text{SEI} [\mathrm{m}^3 \;mol^{-1}]$ & 9.59e-5 & 9.59e-5 & 9.59e-5\\
    $\Omega_\text{pl} [\mathrm{m}^3 \;mol^{-1}]$ & 1.3e-5 & 1.3e-5 & 1.3e-5\\
    
    $\sigma_\text{critical}^+ [MPa]$ & 375 & 375 & 375\\
    $\sigma_\text{critical}^- [MPa]$& 60 & 60 & 60\\

    $T$ [°$C$] & 25 & 25 & 45\\

    $U_\text{SEI} [V]$& 0.4 & 0.4 & 0.4\\
    \bottomrule
    \end{tabular}%
  \caption{SPM and degradation constant parameters}
  \label{tab:SPM parameters}%
\end{table}%

 \bibliographystyle{elsarticle-num} 
 \bibliography{main}





\end{document}